\documentclass[usenatbib]{mn2e}                                                 
\usepackage{graphicx}                                                           
\usepackage{aas_macros}                                                         
\usepackage{amsfonts,amsmath}                                                          
\usepackage{amssymb}
\usepackage{array,hhline}
\usepackage{bm}  
\usepackage{times}                                                              
\usepackage{pifont}
\usepackage{color} 
\begin{document}                                                                
                          

\def\beqn{\begin{equation}}
\def\eeqn{\end{equation}}
\def\Msun{\,\rm M_\odot}                                              
\def\cm{\,{\rm cm}}                                                           
\def\pc{\,{\rm pc}}
\def\kpc{\,{\rm kpc}}                                                           
\def\Gyr{\,{\rm Gyr}}                                                           
\def\Myr{\,{\rm Myr}}                                                           
\def\yr{\,{\rm yr}}                                                           
\def\pyr{\,{\rm yr}^{-1}}
\def\ps{\,{\rm s}^{-1}}
\def\kms{\,{\rm km}\,{\rm s}^{-1}}
\def\gcc{\,{\rm g}\,{\rm cm}^{-3}}
\def\kgcm{\,{\rm kg}\,{\rm m}^{-3}}
\def\K{\,{\rm K}}
\def\keV{\,{\rm keV}}
\def\kgcm2{\,{\rm kg}\,{\rm m}^{-2}}
\def\rc{{r_{\mathrm{c}}}}
\def\rt{{r_{\mathrm{t}}}}
\definecolor{peacock}{rgb}{0.2,0.631,0.788}
\definecolor{deepskyblue4}{rgb}{0,0.408,0.545}
\definecolor{purple}{rgb}{0.502,0,0.502}
\def\wrp{\ding{46}\textbf}
\def\wrpi{\ding{46}\emph}
\def\wrpm{\mathbf}

\def\ud{\mathrm{d}}                                                             

\title[Gas evolution in Galactic globular clusters]{On the evolution of intra-cluster gas within Galactic globular clusters}

\author[W. R. Priestley, M. Ruffert, M. Salaris]
{William Priestley$^{1,3}$\thanks{e-mail: wrp@astro.livjm.ac.uk}, Maximilian Ruffert$^{2}$ and Maurizio Salaris$^{1}$\\
$^1$Astrophysics Research Institute,  Liverpool John Moores University, Twelve Quays House, Egerton Wharf, Birkenhead, Wirral CH41 1LD, United Kingdom\\
$^2$School of Mathematics and Maxwell Institute, University of Edinburgh, King's Buildings, Edinburgh EH9 3JZ, Scotland\\
$^3$National Astronomical Observatory of Japan, 2-21-1 Osawa, Mitaka, Tokyo, 181-8588, Japan 
}
\date{}

\maketitle

\begin{abstract}
It has been known since the 1950's that the observed gas content of Galactic globular clusters (GCs) is $2-3$ orders of magnitude less than the mass lost by stars between Galactic disk crossings. In this work we address the question: What happens to this stellar gas?\\
\indent Using an Eulerian nested grid code, we present 3D simulations to determine how stellar wind material evolves within the GC environment. We expand upon work done in the 70's and move a single-mass King-model GC through the Galactic halo medium, stripping a $10^5\Msun$ GC of its intra-cluster medium but predicting a detectable medium for a $10^6\Msun$ cluster. We find from new multi-mass King model simulations, the first to incorporate empirical mass-loss formulae, that the single-mass King model underestimates the retention of intra-cluster gas in the cluster. Lastly, we present  a simple discretised multi-mass GC model, which yields lower levels of intra-cluster medium compared to the continuous single- and multi-mass King models.\\
\indent Our results show that there is still an issue with the predicted intra-cluster gas content of massive GCs. We conclude that by modelling GC systems more accurately, in particular the stellar structure and description of mass loss, we will be able to work towards resolving this issue and begin to fill in some of the gaps in our understanding of the evolution of globular clusters.
\end{abstract}


\section{Introduction}
\label{sec:intro}

This paper addresses the long-standing problem of why we observe no, or very little, intra-cluster medium (ICM) gas in Galactic globular clusters, when it is understood that giant branch stars are continuously supplying enough material to form an ICM.\\
\indent Galactic globular clusters (GCs) are massive star clusters ($10^4-10^6\Msun$) of approximately $13\Gyr$ in age and are distributed roughly spherically about the Milky Way centre with distances ranging from several to a few tens of $\kpc$ and some as far as $100\kpc$. They orbit the galaxy on timescales of a few $10^8\yr$ \citep{Odenkirchen_et_al97} and will cross the Galactic disk twice per orbit, removing ICM material from the cluster potential each time. The more distant clusters have orbits of a few $\Gyr$ with some perhaps never crossing the Disk\footnote{The Disk and Halo, capitalised, refer to components of the Milky Way.}. There is a wealth of observational evidence for post-main sequence mass loss in GCs from dust tracers \citep[][]{Origlia_et_al97,RamdaniJorissen01,Origlia_et_al02,McDonald_et_al09} and gas tracers \citep{Cohen76,Dupree86,Bates_et_al93,McDonaldVanLoon07}. The winds from these evolved giant branch stars will supply material to the GC ICM on timescales much less than the GC orbital period. Over the whole red giant branch (RGB) phase, stars typically lose $0.2-0.3\Msun$ of gas in the form of cool, slow winds, with velocities around $10-20\kms$, which is below the escape velocities for GCs ($25-75\kms$). Asymptotic giant branch (AGB) stars lose yet another $\sim0.1\Msun$ with velocities of the same order as the RGB winds. Depending on the mass of the GC, it can be expected that $10-100\Msun$ are available to the GC ICM between Galactic disk crossing events \citep{TaylerWood75}.

\subsection{Discrepancy between naive expectations and observations}
\label{subsec:history}

Since the late 1950's, there have been ongoing searches for an ICM that resides within GCs; they have mostly focussed on detecting hydrogen (molecular, atomic, ionised) or dust \citep[see][for a summary]{Roberts88,vanLoon_et_al06}.  These observational studies show two orders of magnitude less gas than expected in any detectable form. Most attempts only place upper limits on the mass of ICM gas with one or two yielding tentative detections. Here, we provide a brief review of the approaches taken towards detecting an ICM.\\
\indent Studies targeting molecular hydrogen (via CO, OH and H$_{\mathrm{2}}$O) yield upper limits of the order of $0.1\Msun$ or no detection \citep{KnappKerr73,Kerr_et_al76,CohenMalkan79,FrailBeasley94,Smith_et_al95,LeonCombes96,vanLoon_et_al06}. Efforts focussed on atomic hydrogen using the $21\cm$ line give upper limits from a few to a few tenths of a solar mass \citep{HeilesHenry66,KerrKnapp72,Knapp_et_al73,ConklinKimble74,Kerr_et_al76,Smith_et_al90,vanLoon_et_al06}. The most sensitive observation to date has been that of \citet{vanLoon_et_al09}, in which no gas is detected in the four clusters they studied. \citet{vanLoon_et_al06} give a tentative detection of $M_{\mathrm{H}}\simeq0.3\Msun$ in M15.\\
\indent Ionised hydrogen searches (via H$\alpha$ and continuum free-free emission) claim upper limits in the range $\sim0.1-1\Msun$ \citep{smith_et_al76,FF77,Knapp_et_al96}. A complementary study by \citet{Freire_et_al01} infers a population of free electrons in the cluster 47 Tucanae from pulsar timing observations, indicating the existence of $0.1\Msun$ of ionised gas within that cluster.\\
\indent In addition to investigations concentrating on hydrogen, there have also been sub-millimetre and infra-red observations focussing on dust \citep{Knapp_et_al95,Hopwood_et_al98,Hopwood_et_al99,Matsunaga_et_al08,Boyer_et_al08,Barmby_et_al09}. These give upper limits on the dust mass of $M_{\mathrm{dust}}\sim10^{-2}-10^{-5}\Msun$. M15 is the only cluster with a detection of dust, where \citet{Evans_et_al03} determine $M_{\mathrm{dust}}\sim5\times10^{-4}\Msun$ and \citet{Boyer_et_al06} find $M_{\mathrm{dust}}\sim9\times10^{-4}\Msun$.\\
\indent Clearly, some mechanism is at work for removing ICM gas from the GC environment.

\subsection{Two theoretical avenues for resolution}
\label{subsec:ideas}

Theoretical investigations into removing gas from GCs can be roughly split into two approaches: mechanisms intrinsic to the GC and those arising from the GC environment. The following is an overview of the theoretical picture to date.\\
\indent Initial investigations looking at intrinsic mechanisms solved steady-state flow equations in one dimension \citep*{Burke68,ScottRose75,FF77}. \citet*{FF77} advanced the work of  \citet{Burke68} and \citet*{ScottRose75} and included energy injection by stellar winds and radiative cooling in the ICM gas. The solution of the steady-state equations require larger RGB wind velocities than are still currently observed. The conclusions warrant further observational efforts and provided motivation for time-dependent simulations of GC ICM gas. The first hydrodynamical simulations \citep*{VF77} traced the evolution of ICM gas in a $10^6\Msun$ GC using RGB wind velocities not permitted by steady-state models ($v_{\mathrm{{\small RGB}\,wind}}\lesssim118\kms$). These lower RGB wind simulations predict that ICM gas will sink to the centre of the cluster and cool. \citet{Vandenberg78} included heating of the ICM due to a UV radiation field from hot horizontal branch stars. This lowered the steady-state wind model threshold RGB wind velocity from $v_{\mathrm{{\small RGB}\,wind}}\simeq118\kms$ to  $v_{\mathrm{{\small RGB}\,wind}}\simeq95\kms$, again providing support for further observational efforts. \citet{Vandenberg78} notes that not all GCs possess hot HB stars. \citet*{ColemanWorden77} included the flaring activity of M-dwarf stars to produce an outflow of ICM gas. The caveat however, is that the flaring properties, numbers and distribution of M-dwarfs within GCs are highly uncertain. Furthermore, modelling the M-dwarf flaring interaction with ICM gas requires additional assumptions. \citet*{ScottDurisen78} produced steady-state models of ICM outflows driven by novae, assuming radiative cooling to be unimportant. However, there are several uncertain factors and novae in GCs are much less common than previously expected \citep{Dobrotka_et_al06,MaccaroneKnigge07} with an estimated rate of $4\times10^{-3}\pyr$ \citep*{BodeEvans08}. Furthermore, novae have axisymmetric outflows \citep{OBrien_et_al06} and will be less efficient at removing gas from the cluster potential.\\
\indent Extraneous to simulations, there are several studies that have investigated potential gas removal mechanisms that may be active within GCs. \citet{Spergel91} looked towards energy liberated from pulsar winds whilst \citet*{YokooFukue92} did the same for X-ray bursters. These letters are brief and only assess the energy requirements for lifting gas from a GC potential well. They contain no detailed investigation as to how energy from these objects is transferred to the ICM gas. \citet{Umbreit_et_al08} propose stellar collisions as a method for removing gas in M15 and \citet{Dupree_et_al09} suggest that low-metallicity RGB winds may be intrinsically fast enough to leave the cluster potential.\\
\indent Extrinsic mechanisms influencing ICM evolution generally concern the effect of the GC moving through the Galactic halo medium \citep*{FrankGisler76}. Such interactions cannot be included in 1D simulations or steady-state models. These authors used an analytical equation to assess the Galactic halo properties required in order to strip a massive GC of ICM gas. They calculated a density roughly an order of magnitude higher than that observed in the Halo and the halos of other galaxies \citep*{Spitzer56,Silk74,FukugitaPeebles06}. To date, no numerical simulations into the Galactic halo-GC ICM interaction have been performed.\\
\indent The questions remain: what happens to gas from giant branch stars and what causes the low ICM content in GCs despite this mass loss? This paper presents the first 3D hydrodynamical simulations to trace the evolution of gas within the globular cluster environment. We model the GC's motion through the Galactic halo and study the effect it has on the ICM gas evolution (which is not possible in spherical symmetry). We provide a more realistic description of the GC stellar population via a multi-mass King model and apply empirical mass-loss laws to describe mass loss within the cluster (from RGB and AGB stars). Using the multi-mass and King models, we also compare the consequence of altering the radial distribution of the stellar mass loss within the GC. We further refine the multi-mass GC into a discrete population of point masses orbiting within the GC potential.\\
\indent This paper is organised as follows. We justify the methods used together with a description of our GC model and present the validation of our hydrodynamics code against \citet*{VF77} in Section \ref{sec:numerical_methods}. In Section \ref{sec:halo_motion}, we discuss the establishment of the GC's motion through the Galactic halo medium. We justify and describe the use of a multi-mass King model with empirical mass-loss laws in Section \ref{sec:MMGC} and continue to examine the effect of discretising the multi-mass model in Section \ref{sec:discrete_stars}. We conclude the paper in Section \ref{sec:Summary}, with a brief summary and discussion of further work that can be done with this kind of investigation.

\section{Numerical methods and validation}
\label{sec:numerical_methods}

\subsection{Hydrodynamics, initial conditions and free parameters}
\label{subsec:method_description}

GCs are essentially spherically symmetric systems where the stellar structure is well represented by 1D models. However, at least a 2D simulation is required in order to simulate the interaction of the ICM gas with the Galactic halo. For the case of a discrete stellar population, a fully 3D simulation is essential. We take a 3D Eulerian hydrodynamics code (derived from {\small PROMETHEUS}) that employs the piecewise parabolic method (PPM) \citep*{ColellaWoodward84} and utilises a fixed nested grid architecture \citep[see][for details]{Ruffert92}. The density scale of the stellar population in GCs spans many orders of magnitude, from the core to the tidal radius. Therefore, nested grids allow us to resolve the dense central regions and cover the whole cluster with minimal computational cost.\\
\indent We follow the same procedure as previous studies in modifying this code \citep{FF77,VF77}. We employ the isotropic (i.e. non-rotating) King model \citep{King66} in order to represent the stellar structure of a globular cluster in our simulation volume. The King model assumes a single homogeneous stellar population (i.e. equal-mass stars), described by a finite isothermal sphere. The distribution function (DF) is a Maxwellian one minus a constant \citep[Equ. (\ref{equ:KingDF}); see][for details]{King66}
\begin{equation}
f\left(r,v_{\ast}\right)\,=\,\kappa {\mathrm{e}}^{-\frac{\left(V-V_\circ\right)}{\sigma^2}}\left({\mathrm{e}}^{\frac{-v_{\ast}^2}{2\sigma^2}}-{\mathrm{e}}^{\frac{-v_{\ast\,{\mathrm{e}}}^{2}}{2\sigma^2}}\right)\;,
\label{equ:KingDF}
\end{equation}
where $\kappa$ is a dimensional constant, $V$ is the local gravitational potential (where $V_{\circ}$ denotes the central value), $\sigma$ is the stellar velocity dispersion and $v_{\ast}$ is the local stellar velocity (where $v_{\ast\,{\mathrm{e}}}$ denotes the local escape velocity). The integration of this DF via the Poisson equation produces a stellar density profile that successfully reproduces the surface brightness profiles of many Galactic GCs (i.e. a flattened core and a cut-off tidal radius where the density falls to zero). In addition to the stellar density profile, the integration of the DF also yields the radial profiles of the gravitational potential and mean squared stellar velocity. The purpose of this stellar distribution is to inject gas into the inter-cluster medium. Stellar mass loss is typically prescribed via the specific mass-loss rate, $\alpha$, that is treated as a free parameter. The product of $\alpha$ and the local stellar density, $\rho_\ast$, gives the local stellar mass-loss rate per unit volume,
\begin{equation}
\dot{\rho}\,=\,\alpha\,\rho_{\ast}\;.
\label{equ:mass_loss_rate}
\end{equation}
The GC model is placed centrally in the simulation volume and implemented by prescribing each cell with a local stellar density, gravitational potential and mean squared stellar velocity, according to its radial position. The hydrodynamic models are advanced in time explicitly and observe the following sequence of steps: the stellar wind material is first injected onto the grid, conserving local mass,
\begin{equation}
\rho_{\mathrm{n+1}}\,=\,\rho_{\mathrm{n}}\, +\, \rho_{\mathrm{inj}}\;,
\label{equ:Kmass_inj}
\end{equation}
momentum,
\begin{equation}
\begin{split}
\rho_{\mathrm{n+1}}\,\bm{v}_{\mathrm{n+1}}\,=&\,\rho_{\mathrm{inj}}\left(\bm{v}_{\mathrm{\ast}}+\bm{v}_{\rm wind}\right)+\rho_{\mathrm{n}}\,\bm{v}_{\mathrm{n}}\\
=&\,\rho_{\mathrm{n}}\,\bm{v}_{\mathrm{n}}\;,
\label{equ:Kmomentum_inj}
\end{split}
\end{equation}
and energy,
\begin{equation}
\rho_{\mathrm{n+1}}\,e_{\mathrm{n+1}}\,=\,e_{\mathrm{n}}\rho_{\mathrm{n}}+U_{\alpha}\rho_{\mathrm{inj}}-\rho_{\mathrm{n+1}}\,\Re\;,
\label{equ:Kenergy_inj}
\end{equation}
accordingly. The momentum term in Equation (\ref{equ:Kmomentum_inj}) for the injected gas vanishes since we assume a spherically symmetric stellar wind and, for an isotropic stellar velocity field, $\bm{v}_\ast=0$ when integrated over all velocity vectors for the `stars' at a given point. After conserving mass, momentum and energy, radiative cooling ($\Re$) is applied before advancing the grid hydrodynamically. The subscripts $\mathrm{n}$ and $\mathrm{n+1}$ refer to the current model and the one being advanced in time by $\Delta t$, respectively, whilst $\mathrm{inj}$ and $\ast$ refer to injected gas and stellar properties, respectively. We adopt the same notation as \citet{VF77} where $\rho$, $\bm{v}$ and $e$ are the ICM gas density, velocity and specific energy, respectively. $U_\alpha$ is the total injected specific energy and contains only the specific kinetic energy injected from the expanding stellar wind, $\beta\left(=\tfrac{1}{2}v_{\mathrm{wind}}^2\right)$, and the change in bulk specific kinetic energy as the injected wind comes to rest relative to the flow of the ambient ICM (i.e. after momentum conservation). VF77 ignore the contribution to $U_\alpha$ from the thermal energy of the stellar winds and so it is not included in our tests. $U_\alpha$ is given by
\begin{equation}
\begin{split}
U_{\alpha}\,=&\,\beta+\frac{1}{2}\left(\bm{v}_\ast-\bm{v}_{\mathrm{n+1}}\right)^2\\
=&\,\beta+\frac{1}{2}\left\langle v_{\ast}^{2}\right\rangle+\frac{1}{2}v_{\mathrm{n+1}}^{2}\;,
\end{split}
\end{equation}
where the cross term, $\bm{v}_{\ast}.\bm{v}_{\mathrm{n+1}}$ from expanding the quadratic disappears due to the isotropic stellar velocity field. $\left\langle v_{\ast}^{2}\right\rangle$ is the local (i.e. dependent on radial position) mean squared stellar velocity and is derived from equation (31) in \citet{King66}.
\begin{equation}
\left<v_{\ast}^{2}\right>\,=\,\frac{4}{3}G\pi {r_{c}}^{2}\rho_{\ast\circ}\left(1-\frac{\frac{2}{5}W^{\frac{5}{2}}}{{\mathrm{e}}^{W}\int_0^W{\mathrm{e}}^{-\eta}\eta^{\frac{3}{2}}d\eta}\right)
\label{equ:msv}
\end{equation}
Equation (\ref{equ:msv}) appears, though with a typographical error, as equation (5) in \citet{FF77}. $W$ is the dimensionless gravitational potential and relates to the gravitational potential via
\begin{equation}
W\,=\,-\frac{V}{\sigma^2}\;
\label{equ:VWrelation}
\end{equation}
and $\eta$ is the dimensionless stellar specific kinetic energy, denoted by
\begin{equation}
\eta\,=\,\frac{v_{\ast}^2}{2\sigma^2}\;.
\label{equ:VEtarelation}
\end{equation}
The denominator in Equation (\ref{equ:msv}) can be integrated by parts to produce a power law series in $W$,
\begin{equation}
\begin{split}
&\sum_{n=3}^\infty\left[\left(\prod_{m=3}^n\frac{2}{2m-1}\right)W^{\frac{2n-1}{2}}\right]\\
&\mathrm{i.e.}\;\frac{2}{5}W^{\frac{5}{2}}+\frac{2}{5}\frac{2}{7}W^{\frac{7}{2}}+\frac{2}{5}\frac{2}{7}\frac{2}{9}W^{\frac{9}{2}}+\cdots\;.
\end{split}
\label{equ:W_pow_law}
\end{equation}
\\
\indent We employ the same King model parameters as \citet*{VF77} (hereafter VF77) to produce a $10^6\Msun$ GC. The parameters adopted are: core radius $\rc=0.5\pc$, tidal radius $\rt=23.9\pc$, central stellar density $\rho_\circ=1.738\times10^{-17}\gcc$ and central dimensionless gravitational potential $W_\circ\,=\,7.5$. We adopt the same metallicity ($X=0.7, Y=0.298, Z=0.002$) and radiative cooling of the ICM as well. The treatment of the radiative cooling is described in \citet*{FF77} and is an analytical equation (their equations (11), (12), (13) and (16)) that approximates a cooling function that models free-free emission, the collisional excitation of permitted lines as well as forbidden and semi-forbidden lines \citep{CoxTucker69,CoxDaltabuit71}. The specific mass-loss rate (for RGB winds) takes the value $\alpha=4\times10^{-19}\ps$. VF77 perform three simulations to probe the effect of the stellar wind velocity on the ICM evolution in GCs. The adopted velocities are $50$, $100$ and $150\kms$; faster than that expected for RGB winds. Stellar wind velocities are a source of energy, injected into the ICM via the specific wind kinetic energy, $\beta$. Our validation simulations are performed on seven nested grids, each a $64^3$ Cartesian mesh. The main (i.e. largest) simulation volume has a side length that is four times the tidal radius. Seven nested grids enables data to be produced at the same length scales in the core as VF77. The initial conditions are a stationary Galactic halo medium with a uniform density of $\rho_{\mathrm{H}}\,=\,10^{-27}\gcc$ at a temperature of $T_{\mathrm{H}}\,=\,10^{5.5}\K$.  We do not attempt to confirm VF77's low pressure Halo simulations where $T_{\mathrm{H}}\,=\,10^{2}\K$.

\subsection{Comparison of our results with those of \citet*{VF77}}
\label{subsec:validateRslts}

We validate our hydrodynamics code with the reproduction of the 1D simulations of VF77 in 3D. We perform the same three simulations for a $10^6\Msun$ GC, as presented in VF77, on our simulation volume. In this section, we split the analysis of the results into two discussions: the unsteady cluster wind simulations (i.e. $\beta$ corresponding to $50\kms$ and $100\kms$) and the steady-state cluster wind simulation ($\beta$ analogous to $150\kms$).\\
\begin{figure}
\begin{center}
\includegraphics[width=0.5\textwidth,angle=0]{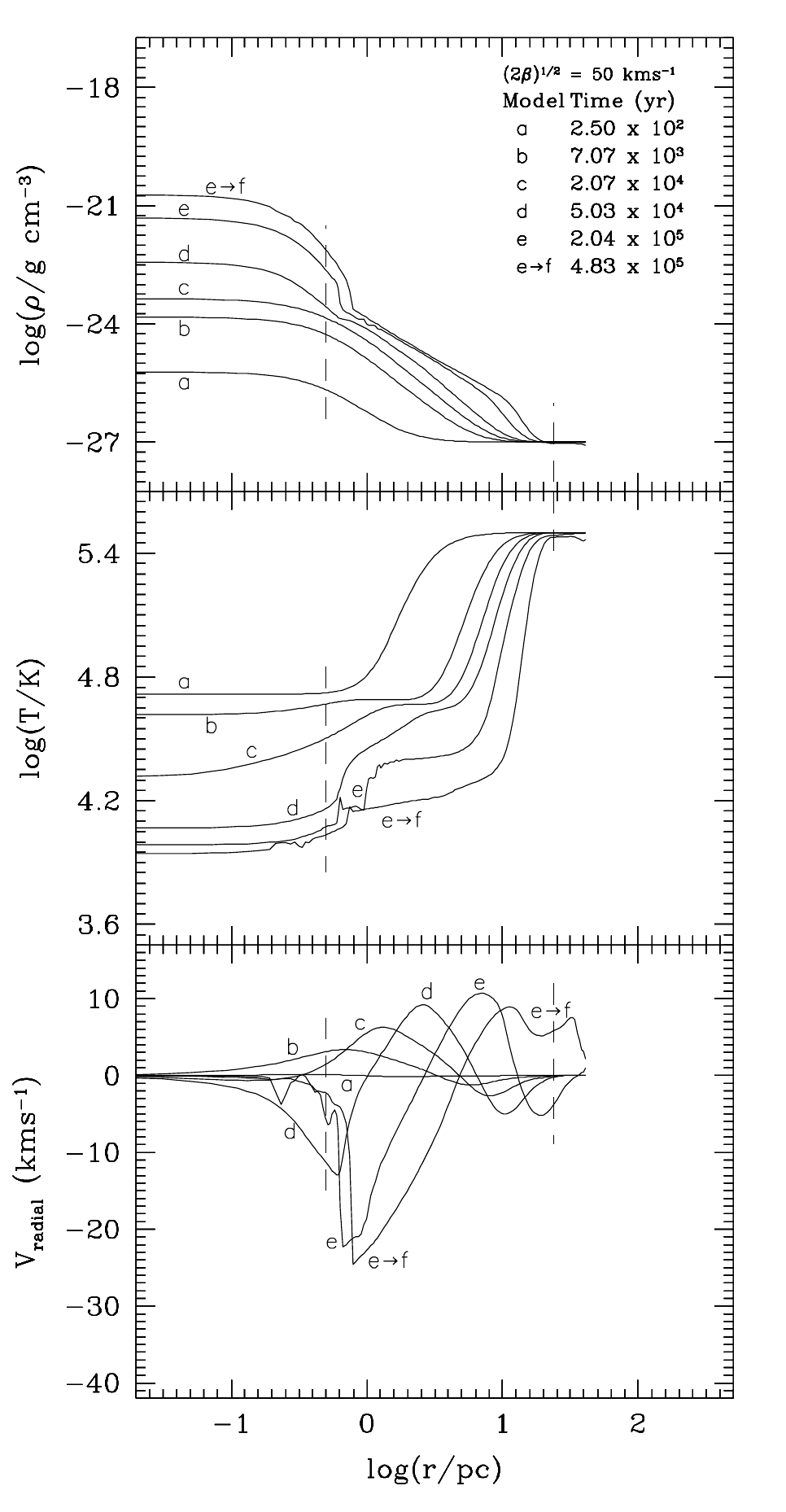}
\caption{A selection of unsteady wind models from our simulations of a $10^6\Msun$ GC with an assumed $v_{\mathrm{wind}}=50\kms$. Model \textsf{e}$\mathsf{\rightarrow}$\textsf{f} shows a model between those of \textsf{E} and \textsf{F} in VF77's fig. 5. The inner and outer dashed vertical lines show the position of the core radius and tidal radius, respectively.\label{fig:50kmsvalidation}}
\end{center}
\end{figure}
\begin{figure}
\begin{center}
\includegraphics[width=0.5\textwidth,angle=0]{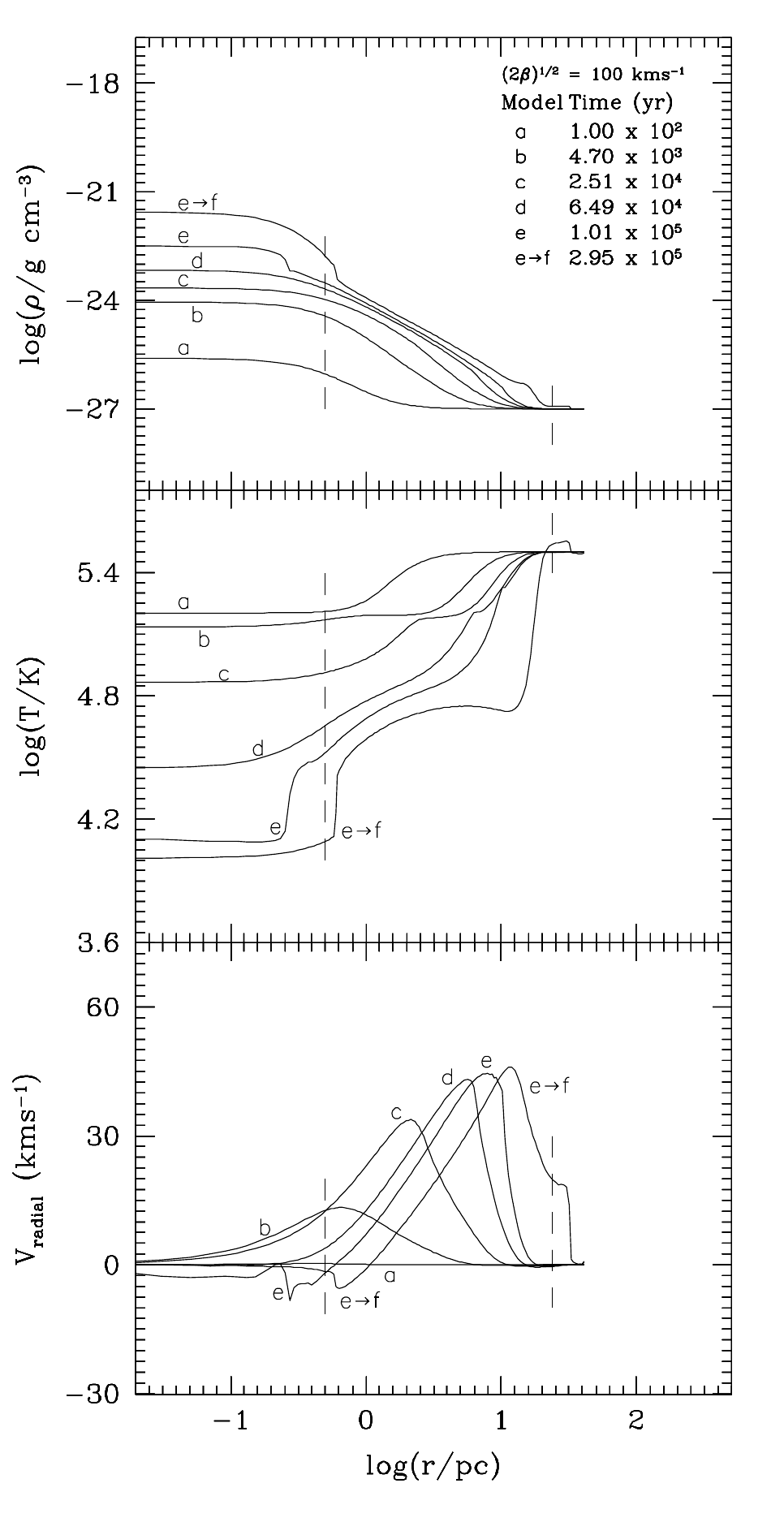}
\caption{A selection of unsteady wind models from our simulations of a $10^6\Msun$ GC with an assumed $v_{\mathrm{wind}}=100\kms$. Model \textsf{e}$\mathsf{\rightarrow}$\textsf{f} shows a model between those of \textsf{E} and \textsf{F} in VF77's fig. 3. The inner and outer dashed vertical lines show the position of the core radius and tidal radius, respectively.\label{fig:100kmsvalidation}}
\end{center}
\end{figure}
\indent In order to save computing time, VF77 begin their models after some time $\Delta t$, performing analytic calculations to determine the initial density, temperature and velocity profile for the gas on their 1D grid. This neglects the effects of cooling in the initial model. Since VF77 do not provide the value of $\Delta t$, we start our simulations from $t\,=\,0$.

\subsubsection{Unsteady cluster wind simulations}
\label{subsubsec:50&150kms_wind}

VF77's results show that the $50\kms$ and $100\kms$ RGB wind velocities result in gas failing to escape the cluster potential, leading to the growth of the ICM within the GC core. They present several models from different times in their simulations (see fig. 5 and fig. 3, respectively, of VF77). A selection of the corresponding models from our simulations are presented in Fig. \ref{fig:50kmsvalidation} and Fig. \ref{fig:100kmsvalidation}, respectively. Due to limited computer resources at the time of validation, we only simulate the minimum number of models required to perform a detailed comparison with VF77. As can be seen in Fig. \ref{fig:50kmsvalidation} and  Fig. \ref{fig:100kmsvalidation}, the main morphological features of the VF77 models are reproduced by our code. The gas density in the central regions increases, forming a bound sphere which cools over time. Gas external to this central reservoir falls onto the surface of the spherical region except at larger radii where gas is seen to move outwards from the cluster. Unsurprisingly, a more detailed comparison with VF77 yields some differences in the gas properties between equivalent models.\\
\indent For the $50\kms$ simulation, we compare the gas properties of some notable features from our model \textsf{e} and VF77's model \textsf{E} (their fig. 5) in Table \ref{tab:50kmsresultscompare}.
\begin{table}\centering
  \setlength\extrarowheight{2pt}
  \caption{Summary of results for the $50\kms$ wind simulation at $0.2\Myr$.}
  \label{tab:50kmsresultscompare}
  \begin{tabular}{ccc}
    \\
    \hline
    ICM properties at $0.2\Myr$ & this work & VF77\\
    \hhline{===}
    $\mathrm{log}\left(\rho_\circ/\gcc\right)$ & $-21.26$ & $-20.5$\\
    $\mathrm{log}\left(T_\circ/\K\right)$ & $3.99$ & $3.93$\\
    $v_{\mathrm{radial,\,max}}/\kms$ & $11$ & $5$\\
    $v_{\mathrm{radial,\,min}}/\kms$ & $-22$ & $-30$\\
    $\mathrm{log}\left(r_{\mathrm{reservoir}}/\pc\right)$ & $-0.175$ & $-0.51$\\
    \hline
    \\
  \end{tabular}
\end{table}
We find our central density to be almost a sixth of VF77's and the extent of the central reservoir of gas is twice the size. The central temperature is $15\%$ higher, likely the result of a lower cooling rate in the lower-density reservoir. The peak inflow velocity from VF77, shows a higher inflow velocity due to gas falling onto a more compact core. The outflow velocity feature coincides at the same radii, yet our results have a velocity that is just over twice that of VF77.\\
\begin{table}\centering
  \setlength\extrarowheight{2pt}
  \caption{Summary of results for the $100\kms$ wind simulation at $0.1\Myr$.}
  \label{tab:100kmsresultscompare}
  \begin{tabular}{ccc}
    \\
    \hline
    ICM properties at $0.1\Myr$ & this work & VF77\\
    \hhline{===}
    $\mathrm{log}\left(\rho_\circ/\gcc\right)$ & $-22.5$ & $-21.75$\\
    $\mathrm{log}\left(T_\circ/\K\right)$ & $4.1$ & $4.025$\\
    $v_{\mathrm{radial,\,max}}/\kms$ & $44.75$ & $47.5$\\
    $v_{\mathrm{radial,\,min}}/\kms$ & $-8.75$ & $-17.5$\\
    $\mathrm{log}\left(r_{\mathrm{reservoir}}/\pc\right)$ & $-0.575$ & $-0.7$\\
    \hline
    \\
  \end{tabular}
\end{table}
\indent For the $100\kms$ wind simulations, we compare model \textsf{e} with model \textsf{E} of VF77 (their fig. 3). In Table \ref{tab:100kmsresultscompare}, the central density is nearly a sixth the value in VF77 and the central temperature is about $19\%$ higher. The radial extent of our core reservoir of gas is about $33\%$ larger. The peak inflow velocity is half that of VF77, and occurs at a slightly larger radius. The outflow peak velocity is just $6\%$ less than VF77's.\\
\indent VF77 do not publish their ICM gas energy data, therefore, the temperature data presented in Tables \ref{tab:50kmsresultscompare} and \ref{tab:100kmsresultscompare} are for reference purposes only. We warn the reader that it is inadvisable to directly compare the gas temperatures because, unlike energy, it is not solved for in the conservative form of the hydrodynamics equations.\\
\indent Aside from the initial models used in the corresponding simulations, there are other systematic differences leading to a divergence in the solutions. The implicit method employed by VF77 and our explicit scheme are fundamentally different and, naturally, will not lead to exactly the same numerical solution. Furthermore, a 1D simulation requires a central boundary condition. VF77 apply the commonly adopted condition that $v=0$ at $r=0$, therefore, gas approaching the centre must become stationary and cannot flow freely around this point. An excellent example of this restriction can be seen in model \textsf{d} of Fig. \ref{fig:100kmsvalidation}, where the gas has a velocity of $-2.5\kms$ at very small radii. In contrast, curve \textsf{D} of fig. 3 (VF77) shows this negative velocity approaching zero at smaller radii. We argue that this boundary condition artificially traps gas at smaller radii leading to a more condensed and cooler reservoir of ICM gas. With these in mind, and considering that the initial density is five to six orders of magnitude lower, the $\sim0.75{\,{\rm dex}}$ difference in the central density (Tables \ref{tab:50kmsresultscompare} and \ref{tab:100kmsresultscompare}) is not as alarming as it first seems. Considering these systematic differences, it is encouraging that the morphology is faithfully reproduced by our code.

\subsubsection{Steady-state cluster wind simulation}
\label{subsubsec:150kms_wind}

\begin{figure}
\begin{center}
\includegraphics[width=0.5\textwidth,angle=0]{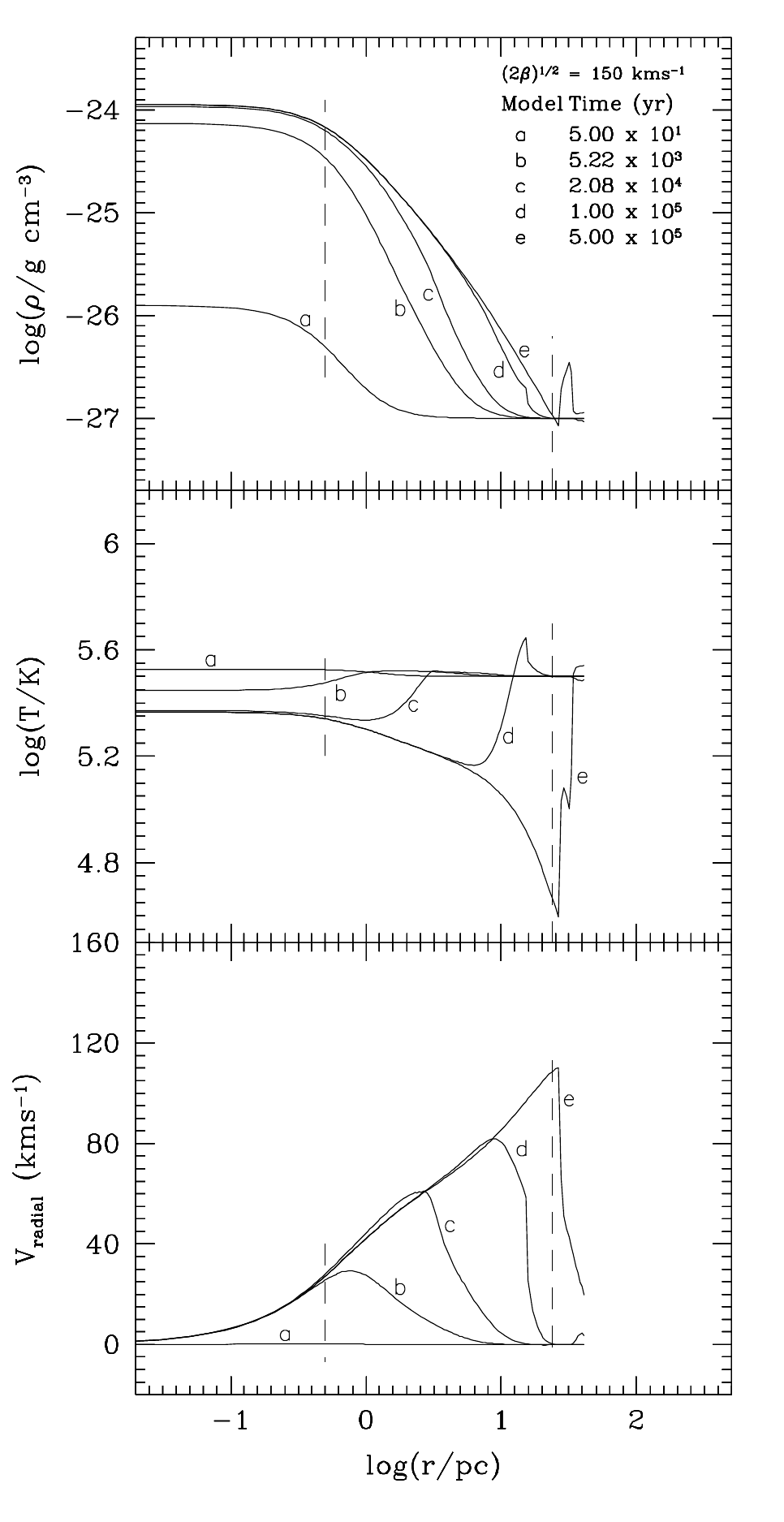}
\caption{A selection of the steady-state wind models from our simulations of  a $10^6\Msun$ GC with an assumed $v_{\mathrm{wind}}=150\kms$. The inner and outer dashed vertical lines show the position of the core radius and tidal radius, respectively.\label{fig:150kmsvalidation}}
\end{center}
\end{figure}
The $150\kms$ RGB wind simulation leads to a steady-state outflow of ICM gas from the globular cluster (see VF77, fig. 1). VF77 compare their GC wind model (curve \textsf{E}) with the steady-state solution of \citet{FF77} (the dotted line in their fig. 1) and find them to be in excellent agreement. Since the evolution of the ICM leads to a steady-state solution, the central boundary condition exerts little influence on the gas dynamics. Furthermore, cooling does not dominate the flow for this RGB wind velocity, reducing the number of factors to consider compared to the unsteady GC wind case. Consequently, we therefore expect an overall closer agreement with VF77.\\
\begin{table}\centering
  \setlength\extrarowheight{2pt}
  \caption{Summary of features in the steady-state model at $0.5\Myr$ for the $150\kms$ wind simulation.}
  \label{tab:150kmsresultscompare}
  \begin{tabular}{ccc}
    \\
    \hline
    ICM properties at $0.5\Myr$ & this work & VF77\\
    \hhline{===}
    $\mathrm{log}\left(\rho_\circ/\gcc\right)$ & $-23.95$ & $-24.01$\\
    $\mathrm{log}\left(T_\circ/\K\right)$ & $5.37$ & $5.48$\\
    $v_{\mathrm{radial}}\left(\rt\right)/\kms$ & $109$ & $117$\\
    $M_{\mathrm{ICM}}/\Msun$ & $2.83$ & $2.59$\\
    \hline
    \\
  \end{tabular}
\end{table}
\indent Fig. \ref{fig:150kmsvalidation} shows models from our simulation and is the equivalent of fig 1. in VF77. Since a GC wind is established within relatively short timescales, we were able to reproduce the same models presented in VF77. Encouragingly, the two sets of models are remarkably similar. However, the temperature profile of models \textsf{a}, \textsf{b} and \textsf{c}, show a striking disparity, which is less so in later models. Our models show a lower temperature, at low and intermediate radii, compared to VF77. This difference, is caused by the absence of cooling in VF77's method for determining their initial model. To illustrate this, we put $\Delta T=50\yr$ (i.e. the time to reach model \textsf{a}) into equations (14) and (15) of VF77. The corresponding central density and temperature are $\log\left(\rho_\circ/\gcc\right)=-25.92$ and $\log\left(T_\circ/\K\right)=5.73$. The estimated density agrees, to within $5\%$, with our model \textsf{a}, however, the analytical temperature is $58\%$ higher than ours. Compared to VF77's  central gas temperature of $\log\left(T_\circ/\K\right)=5.69$, our model \textsf{a} has a temperature that is $31\%$ lower at $\log\left(T_\circ/\K\right)=5.53$. Table \ref{tab:150kmsresultscompare} gives a quantitative comparison of some notable features in model \textsf{e}, the model showing a steady-state outflow of gas from the GC. The central density we obtain is about $15\%$ higher than VF77, and the central temperature $22\%$ lower. At the tidal radius, our radial gas velocity is about $6\%$ lower than VF77. Due to the denser, cooler central gas parameters, our ICM is $9\%$ more massive. Despite some differences in the early models, the overall morphology at steady state is almost identical, with a quantitative agreement to within $10-20\%$.

\subsubsection{Summary of comparison}
\label{subsubsec:summary}

In summary, we run simulations employing the same GC models, gas physics, Halo gas properties and RGB stellar wind velocities, as presented in \citet{VF77}. We compare the results in order to validate our hydrodynamics code. In all cases, the ICM evolution is qualitatively reproduced, with equivalent models showing few morphological differences. We identify several systematic quantitative differences in the results, which we attribute to several factors: the fundamental difference between an implicit and an explicit code in reaching a solution; a 1D mesh requires a central boundary condition of $v=0$ at $r=0$ that will exert a ``braking'' influence on the gas dynamics in the central regions; VF77's initial models start after some undefined time period $\Delta t$ and ours at $t=0$. In spite of these points, our code reproduces the ICM evolution presented in VF77 remarkably well.\\
\indent To a crude approximation, simulating from $t=0$ can be taken to represent the environment shortly after a catastrophic ICM-removal event.

\section{Moving through the halo environment}
\label{sec:halo_motion}

The GC model in section \ref{subsec:validateRslts} was at rest with respect to the ambient medium. Obviously GCs are not stationary within the Galactic halo but in continuous motion through this hot, tenuous ambient gas \citep[$T\sim10^6\K$, $\rho\sim10^{-27}\gcc$;][]{Spitzer56} at supersonic speeds between $100\kms$ and $300\kms$ \citep{Odenkirchen_et_al97}. \citet*{FrankGisler76} write that their analytical formula is confirmed by the results of \citet{Gisler76}, who undertakes 2D simulations of elliptical galaxies in clusters. There have been no direct simulations investigating a GC's motion within the Galactic halo in the literature to date. In order to determine if such simulations are justified, we use equation (2) from \citet*{FrankGisler76} to work out the Halo density required to continuously strip our model GC of its ICM gas. We roughly estimate the surface density of stars at the centre of the cluster, $\sigma_\circ$, to be $\rho_\circ\rc=134\kgcm2$. We use $v_{\small \rm GC}=200\kms$ as a typical GC velocity, which is the median of the distribution of values presented in \citet{Odenkirchen_et_al97}. The escape velocity for our cluster is $v_{\rm esc}\simeq76\kms$ and $\alpha=4\times10^{-19}\ps$, as in Section \ref{subsec:method_description}. The corresponding Halo density, must be around $10^{-25}\gcc$ in order to remove all gas from the GC; two orders of magnitude higher than the upper Halo value given by \citet{Spitzer56}. Therefore, away from the Disk, we should still expect GCs to build-up some form of ICM.

\subsection{What environmental factors are important?}
\label{subsec:intro_to_tests}

We perform simulations of the GC moving through the ambient Halo medium. This is achieved by following the GC's motion in the GC's frame of reference and choosing a set of boundary conditions that produces a continuous flow of ambient Halo gas past the GC. The GC remains fixed at the centre of the simulation volume. In all simulations, gas flows in the positive $x$ direction of the grid. We run several low-resolution simulations, using three nested grids at a resolution of $ 64^3$, in order to test the influence of several parameters on ICM evolution. These parameters are, the GC velocity, the specific mass-loss rate ($\alpha$) and the density of the Galactic halo medium. A convergence test showed, that using three nested grids rather than seven still gives the same overall results regarding ICM gas physics and morphology, but slightly overestimates the ICM gas content. However, an understanding of the influence of different parameters over longer evolutionary timescales is worth the trade-off in accuracy of the predicted ICM mass. The GC velocity is set to either a median $200\kms$ or an upper value of $300\kms$. The specific mass-loss rate, $\alpha$, takes either one of the two extreme values $4\times10^{-19}\ps$ \citep[determined in][]{Knapp_et_al73} and $4\times10^{-20}\ps$ \citep{TaylerWood75}. Finally, in addition to the assumed $10^{-27}\gcc$, we also give the Halo density an extreme value of $10^{-26}\gcc$ in some simulations. We provide a list of the initial conditions used for each low-resolution simulation in Table \ref{tab:LowResHalo}. The stellar wind velocity is assumed to be $20\kms$ for all simulations.
\begin{table}\centering
  \setlength\extrarowheight{2pt}
  \caption{Initial conditions for the low-resolution simulations of a GC moving through the Galactic halo medium.}
  \label{tab:LowResHalo}
  \begin{tabular}{ccccc}
    \\
    \hline
    Simulation & $M_{\mathrm{GC}}$ & $\rho_{\mathrm{H}}$ & $v_{\rm \small GC}$ & $\alpha$ \\
     & $\left(\Msun\right)$ & $\left(\gcc\right)$ & $\left(\kms\right)$ & $\left(\ps\right)$ \\
    \hhline{=====}
    \textbf{A} & $10^6$ & $10^{-26}$ & $200$ & $4\times10^{-19}$\\
    \textbf{B} & $10^6$ & $10^{-26}$ & $200$ & $4\times10^{-20}$\\
    \textbf{C} & $10^5$ & $10^{-26}$ & $200$ & $4\times10^{-19}$\\
    \textbf{D} & $10^6$ & $10^{-27}$ & $200$ & $4\times10^{-19}$\\
    \textbf{E} & $10^6$ & $10^{-27}$ & $200$ & $4\times10^{-20}$\\
    \textbf{F} & $10^5$ & $10^{-27}$ & $200$ & $4\times10^{-19}$\\
    \textbf{G} & $10^6$ & $10^{-27}$ & $300$ & $4\times10^{-19}$\\
    \hline
    \\
  \end{tabular}
\end{table}

\subsection{ICM-Halo gas interaction}
\label{subsec:HaloRslts}

\begin{figure}
\begin{center}
\includegraphics[width=0.5\textwidth,angle=0]{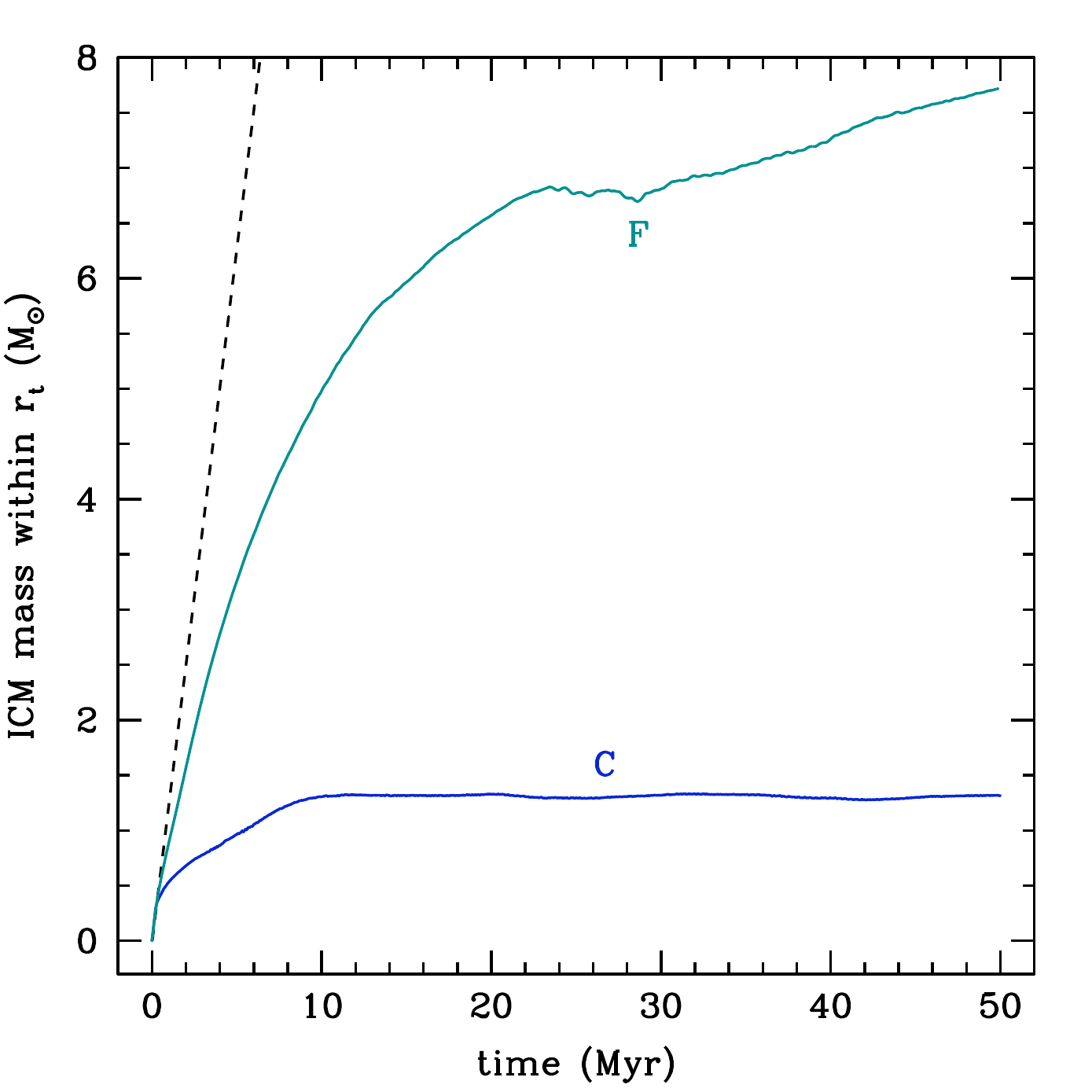}
\caption{The ICM content over time for two $10^5\Msun$ GCs with specific mass-loss rate $\alpha=4\times10^{-19}\ps$, moving through a Galactic halo medium at $200\kms$. Simulation \textbf{C} uses a Halo density of $10^{-26}\gcc$ and \textbf{F} uses $10^{-27}\gcc$. The black dotted line is the injected gas mass against time.\label{fig:CF_compare}}
\end{center}
\end{figure}

We ran simulations for at least $10\Myr$ (i.e. $10\%$ of the time between Disk crossings). We also compare models from different simulations at this time. We first compare the ICM gas mass evolution for the $10^5\Msun$ GC simulations (\textbf{C} and \textbf{F}), which are shown in Fig. \ref{fig:CF_compare}. For the $10^6\Msun$ GC simulations, we address the analysis of the two assumed specific mass-loss rates separately. Fig. \ref{fig:ADG_compare} shows simulations \textbf{A},  \textbf{D} and \textbf{G}, which employ the higher mass-loss rate of $\alpha=4\times10^{-19}\ps$. Fig. \ref{fig:BE_compare} shows simulations \textbf{B} and \textbf{E} using the lower rate of $\alpha=4\times10^{-20}\ps$. We finally compare the relative gas content for all simulations in Fig. \ref{fig:relative_gas_content}, which shows the ICM mass as a fraction of the total injected gas with time.

\subsubsection{``Typical'' mass GC simulations}
\label{subsubsec:halo_1E5}

\begin{figure}
\begin{center}
\includegraphics[width=0.5\textwidth,angle=0]{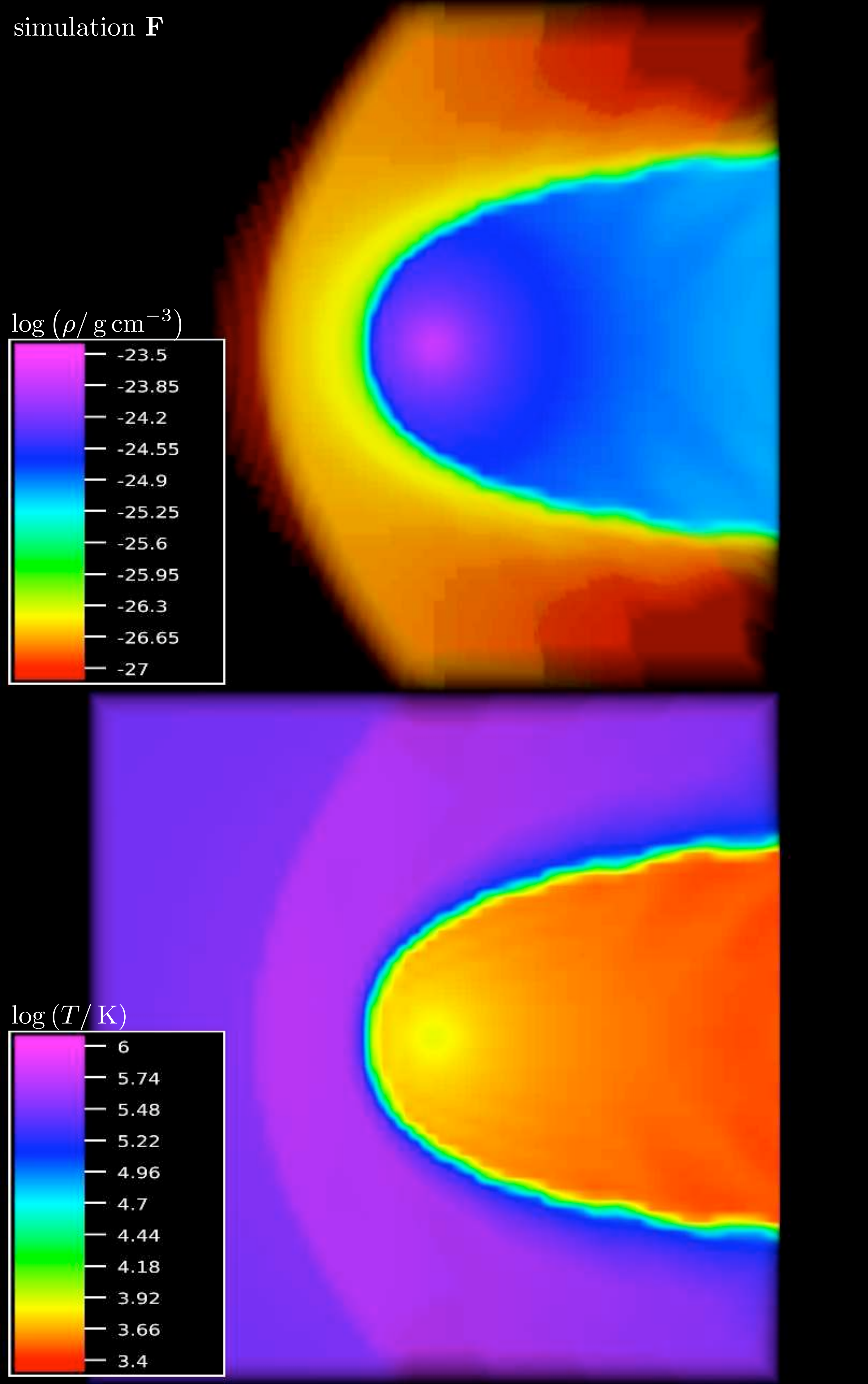}
\caption{Simulation \textbf{F}: A cutaway of a 3D rendered image showing $\log\left(\rho/\gcc\right)$ (upper panel) and $\log\left(T/\K\right)$ (lower panel) of the ICM gas from a $10^5\Msun$ GC simulation at $10\Myr$. These data are taken from the $3^{\mathrm{rd}}$ grid that has a side length of $23.9\pc$.\label{fig:F_lowres_pic}}
\end{center}
\end{figure}

\begin{figure}
\begin{center}
\includegraphics[width=0.5\textwidth,angle=0]{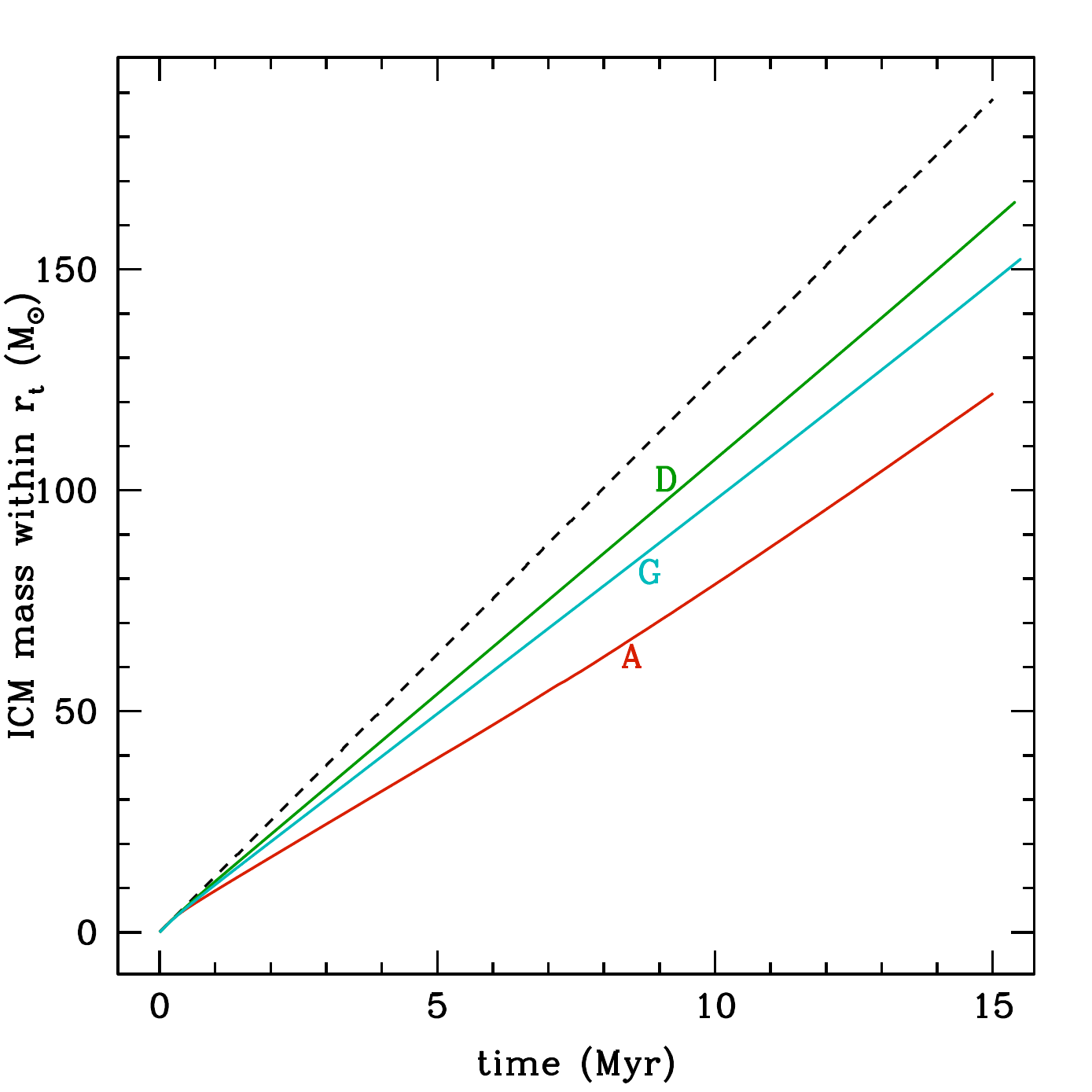}
\caption{The ICM content over time for three $10^6\Msun$ GCs with specific mass-loss rate $\alpha=4\times10^{-19}\ps$, moving through a Galactic halo medium. Simulations \textbf{A} and \textbf{D} move at $200\kms$ through Halo densities of $10^{-26}\gcc$ and $10^{-27}\gcc$, respectively. Simulations \textbf{D} and \textbf{G} move though a Halo density of $10^{-27}\gcc$ with velocities of $200\kms$ and $300\kms$, respectively. The black dotted line is the injected gas mass against time.\label{fig:ADG_compare}}
\end{center}
\end{figure}

\begin{figure}
\begin{center}
\includegraphics[width=0.5\textwidth,angle=0]{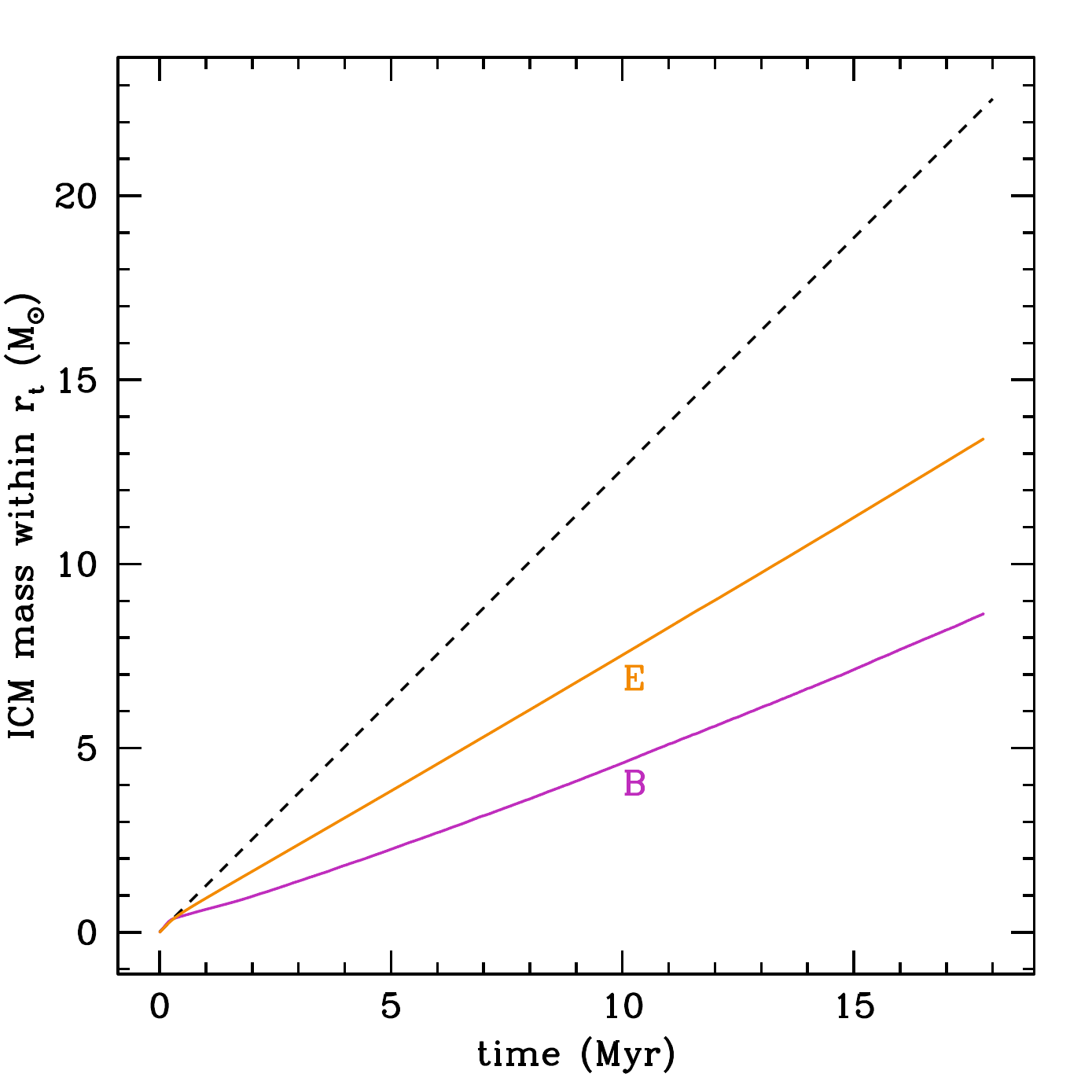}
\caption{The ICM content over time for two $10^6\Msun$ GCs with specific mass-loss rate $\alpha=4\times10^{-20}\ps$, moving through a Galactic halo medium at $200\kms$. Simulation \textbf{B} uses a density of $10^{-26}\gcc$ and \textbf{E} uses $10^{-27}\gcc$. The black dotted line is the injected gas mass against time.\label{fig:BE_compare}}
\end{center}
\end{figure}

The $10^5\Msun$ GC simulations (\textbf{C} and \textbf{F}) represent a ``typical'' GC mass \citep{GnedinOstriker97} and so the results are significant for the majority of the Galactic GCs we observe.\\
\indent Here, we investigate how the Halo density affects the ICM gas evolution. The ICM content within these clusters is strongly influenced by the density of the Galactic halo medium, with the final ICM mass in simulations \textbf{C} and \textbf{F} (Fig. \ref{fig:CF_compare}) differing by nearly a factor of $6$. Simulation \textbf{C} reaches a steady state by about $10\Myr$ and contains (i.e. material within $\rt$) about $1.3\Msun$ of gas for the remainder of the simulation. The ICM mass in simulation \textbf{F} starts to reach a plateau at $\sim24\Myr$ then, after $4\Myr$, continues to increase roughly linearly. The rate at which the gas mass increases is $4\%$ of the total stellar mass-loss rate. At $10\Myr$, \textbf{F} contains $\sim5\Msun$ of ICM gas, which is $40\%$ of the injected  material, i.e. $60\%$ has been stripped away by this point. For \textbf{C}, $90\%$ of the injected material has been removed by $10\Myr$. By $50\Myr$ the total ICM content in \textbf{F} is $7.7\Msun$ ($12.3\%$ of injected material) whilst for \textbf{C}, an impressive $98\%$ of the injected gas has been stripped from the cluster.\\
\indent Because efforts to detect gas within GCs focus primarily on hydrogen, we determine the ${\rm H}$ content of the ICM by multiplying the ICM mass by the hydrogen mass fraction ($\mathrm{X}=0.7$). At $10\Myr$, the $\mathrm{H}$ mass content in simulations \textbf{C} and \textbf{F} are $0.9\Msun$ and $3.5\Msun$, respectively, whilst at $50\Myr$, \textbf{F} contains $5.4\Msun$ of hydrogen (\textbf{C} still contains $\sim0.9\Msun$). However, this is just the gas that resides within the cluster tidal radius. Fig. \ref{fig:F_lowres_pic} shows that the ICM gas predominantly resides in a trailing tail of stripped material.
Only a fraction of this total hydrogen gas resides within the core region. Selecting an arbitrary radius of $2\pc$, the fraction of ICM gas within this region is $15\%$ ($0.13\Msun$) for \textbf{C} and $9\%$ ($0.32\Msun$) for \textbf{F}.\\
\indent In addition to the morphological structure of the ICM, Fig. \ref{fig:F_lowres_pic} also shows the ICM temperature to be lower than $10^4\K$, indicating that hydrogen will be almost entirely neutral; indeed, using equation (11) of \citet{FF77} shows this to be the case. This result suggests that HI, rather than H$_\alpha$ observations will provide more insightful information about the ICM content in GCs.\\
\indent In summary, the predicted gas content of ``typical'' GCs is around the upper limit values given in the literature. However, due to the morphology of the ICM gas, the actual mass of gas that one would determine from observational data will clearly depend on the GC's direction of motion relative to us.\\
\indent In conclusion, ICM gas is continuously stripped from typical clusters and the residual gas content will prove particularly challenging to detect with current radio telescopes. The bow shock temperature of around $10^6\K$ (or $0.08\keV)$, is consistent with the X-ray emission temperatures observed by \citet{Okada_et_al07}.

\subsubsection{High mass GC simulations}
\label{subsubsec:halo_1E6}

\begin{figure}
\begin{center}
\includegraphics[width=0.5\textwidth,angle=0]{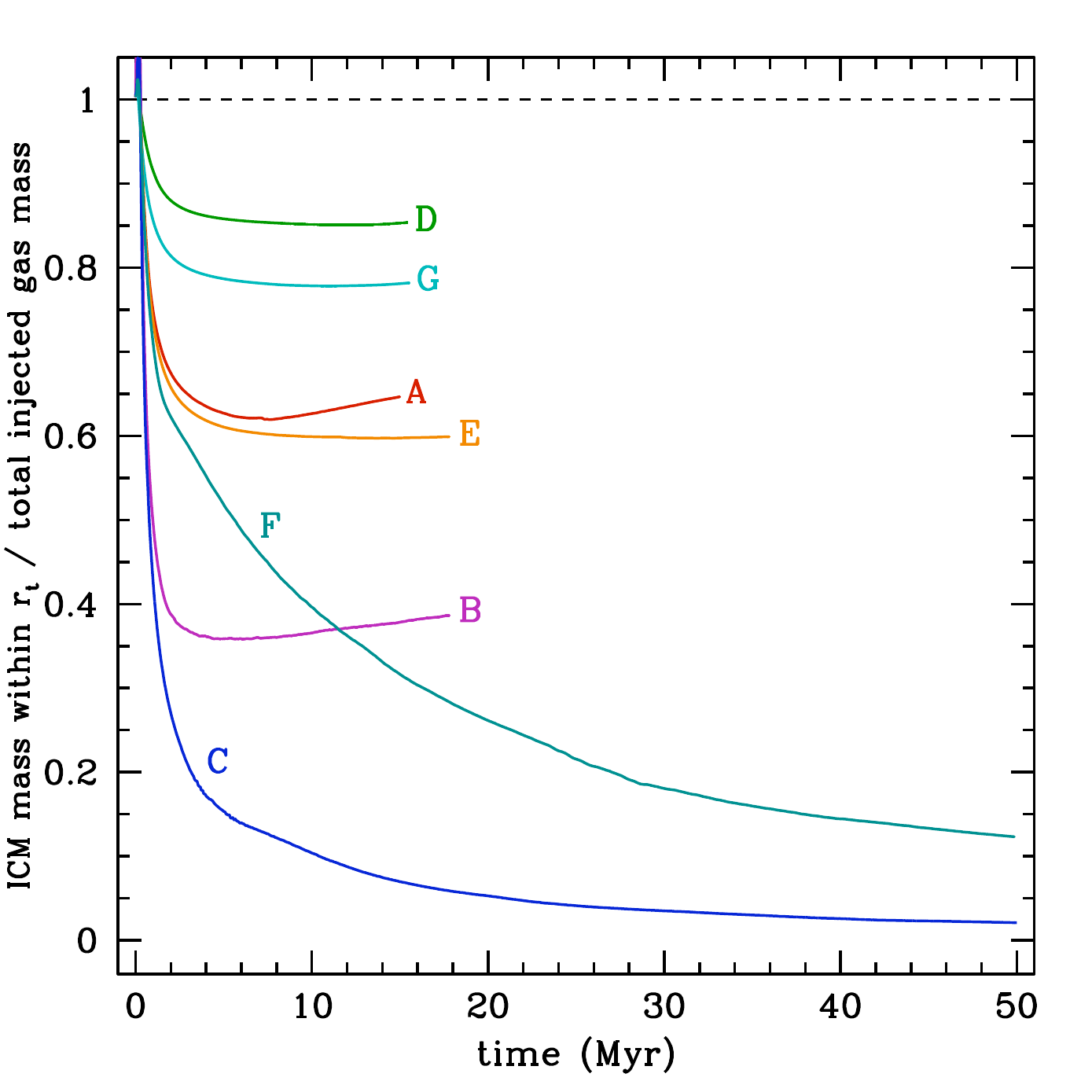}
\caption{The ICM mass within different GC simulations as a fraction of the cumulative injected gas mass against time.\label{fig:relative_gas_content}}
\end{center}
\end{figure}

The $10^6\Msun$ clusters represent the upper tail of the Galactic GC mass distribution. Removing gas from them poses a particular challenge, since these clusters have much deeper gravitational potential wells as well as more stars contributing to the ICM. Unlike the $10^5\Msun$ GC simulations, where the ICM content eventually remains constant with time, the $10^6\Msun$ simulations (\textbf{A}, \textbf{B}, \textbf{D}, \textbf{E} and \textbf{G}) show a continuous increase (Fig. \ref{fig:ADG_compare} and Fig. \ref{fig:BE_compare}). This increase occurs almost linearly, at some fraction of the rate of injected stellar gas. It is clear that this linearity is brought about by a balance between the force due to ram pressure and that due to gravity; where the rate of ram-pressure stripping is less than the injection rate of the stellar material. This is demonstrated in Fig. \ref{fig:relative_gas_content}, where the curves flatten out once this equilibrium is established. This happens on a timescale of about $5\Myr$. The invariability of this equipoise over periods of ${\rm Myr}$ is nicely illustrated by simulations \textbf{D}, \textbf{E} and \textbf{G}, where the curves remain almost parallel with the injected mass of gas. For \textbf{A} and \textbf{B}, the ICM mass relative to the total injected stellar material begins to rise again after a while. This is simply due to the accumulation of the denser upstream Halo medium. All of the $10^6\Msun$ simulations establish a bound ICM, where mass increases over time above the observed upper limits. At $10\Myr$, simulation \textbf{A} contains $78.9\Msun$, \textbf{D}, $107.1\Msun$ and \textbf{G} contains $97.9\Msun$, orders of magnitude above observational limits. Simulations \textbf{B} and \textbf{E}, which adopt the lower value of $\alpha$, contain $4.6\Msun$ and $7.5\Msun$, respectively, by $10\Myr$.\\
\indent It is clear that even under extreme conditions (of Galactic halo density or GC velocity), the environment alone cannot prevent a build-up of bound ICM. The factor that produced the greatest influence on the ICM content is the specific mass-loss rate. Adopting a lower value of $\alpha$, affects the ICM evolution in two ways: firstly, it reduces the amount of gas being injected per unit time, trivially resulting in less gas to be removed. Secondly, for a given ram pressure, a lighter ICM will have a larger fraction lifted out of the GC potential. This latter effect is evident in Fig. \ref{fig:relative_gas_content} from the relative locations of \textbf{D} and \textbf{E}, as well as \textbf{A} and \textbf{B}, where simulations \textbf{D} and \textbf{E} have the same ram pressure but different values of $\alpha$. The same is true for \textbf{A} and \textbf{B}, which experience a higher ram pressure than \textbf{D} and \textbf{E}. The lower $\alpha$ values constitute a $30-40\%$ reduction in the \emph{relative} amount of remaining ICM per unit time. Considering the impact of one's choice of $\alpha$ and the restrictions on varying the environmental parameters, we are led to conclude that modelling the GC correctly and how mass is lost from the stellar population is a key factor in furthering our understanding of the evolution of ICM gas within GCs.\\
\indent Fig. \ref{fig:relative_gas_content} shows that the ICM gas content in $10^5\Msun$ GCs is affected by the Halo density to a greater degree than in the $10^6\Msun$ GCs (simulations \textbf{A} and \textbf{D}). In Fig. \ref{fig:relative_gas_content}, we note that the initial mass of ICM gas compared to the injected stellar mass increases above $1$ (by up to $10\%$). This is due to the congestion of upstream Halo gas interacting with the injected stellar gas at early epochs. This is evident from the location of the peak of this initial increase, which occurs when we expect stripped central Halo-ICM gas to begin crossing $\rt$ ($\sim0.1\Myr$). This also implies that the ICM gas within $\rt$ for the $10^5\Msun$ GCs is refreshed with new material on this timescale.

\section{Multi-mass King model}
\label{sec:MMGC}

We move to a multi-mass model of the GC in order to better describe both the distribution of mass loss as well as the prescription of the mass loss itself within our GC model. To date, the prescribed mass loss has been treated as essentially a free parameter, $\alpha$, where values are guided from observed cluster properties \citep{Knapp_et_al73,TaylerWood75}. This blanket mass-loss rate is applied over the entirety of the GC. This approach is not entirely accurate, as it is the RGB and AGB stars that almost exclusively provide material for the ICM gas. Furthermore, mass loss in these stars, although still not well understood, has been studied extensively. Hence, we can prescribe mass-loss rates according to relevant empirical formulae \citep[][respectively]{Reimers75,VassiliadisWood93} that are derived from stellar properties, not the heterogeneous properties of a GC. The initial investigations presented in Sections \ref{sec:numerical_methods} and \ref{sec:halo_motion} use the King model, which is derived using the assumption that all the stars have the same mass. However, a real GC consists of a stellar population spanning a range of masses, where radial distributions are subject to mass segregation via two-body relaxation. This means that massive giant branch stars are concentrated towards the core compared to an average distribution that is dominated by the more numerous lighter stars. In order to model the AGB and RGB stars as separate populations with different radial distributions and independent mass-loss rates, we require a multi-mass King model. A multi-mass King model is one in which the total mass and overall stellar density distribution is constructed out of several single-mass King models, each one made from a different stellar mass.

\subsection{Implementation of the multi-mass King model}
\label{subsec:MMGC_implement}

\begin{figure}
\begin{center}
\includegraphics[width=0.5\textwidth,angle=0]{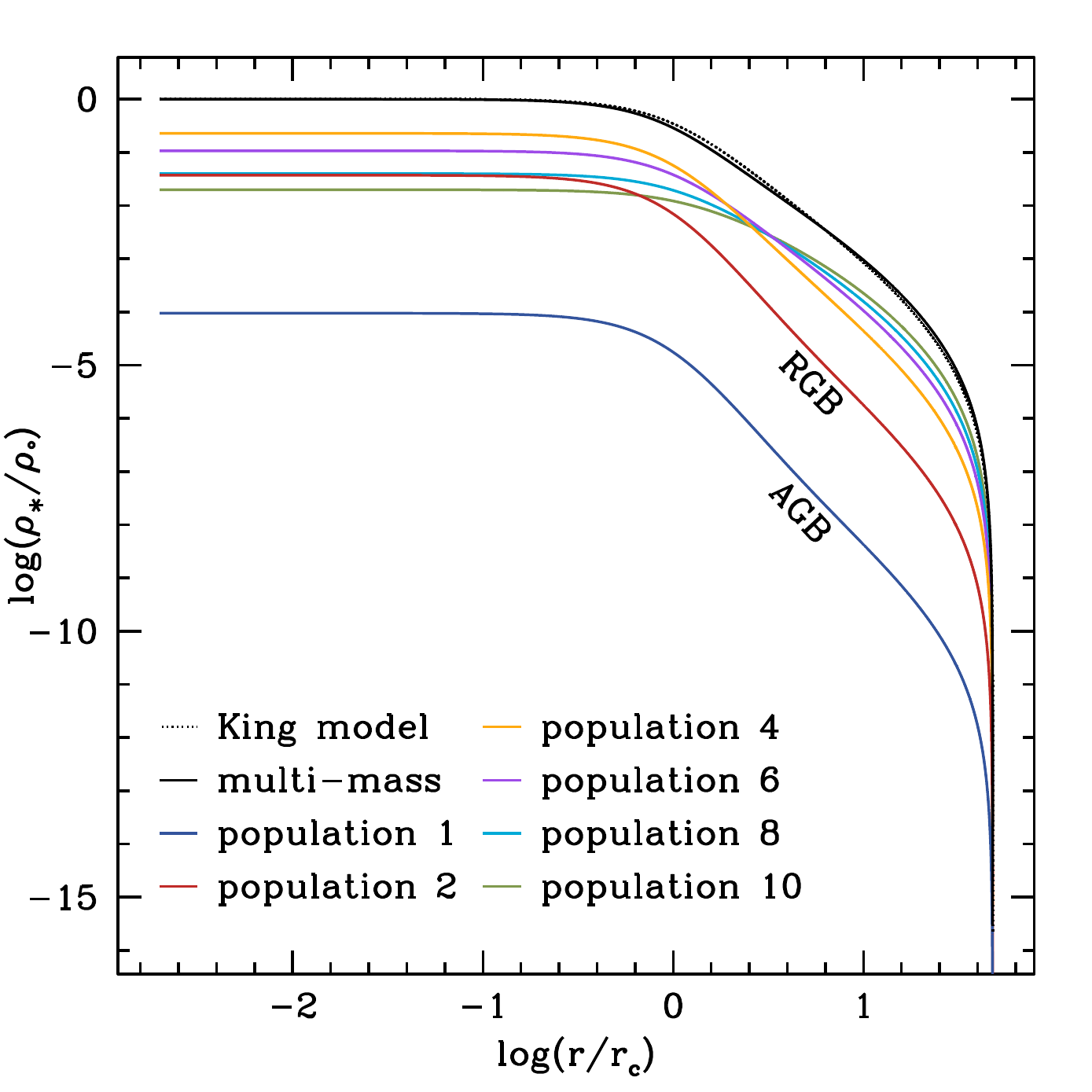}
\caption{The stellar density of a King GC and multi-mass GC model. The values are given as a fraction of the central density, $\rho_\circ$, of the King model. The dotted line is the density profile of the King GC profile and the solid black line is the sum of the component sub-King models, a sample of which are represented by the coloured lines. The AGB and RGB populations are denoted by populations 1 and 2, respectively.\label{fig:MMGCmodel}}
\end{center}
\end{figure}

\begin{figure}
\begin{center}
\includegraphics[width=0.5\textwidth,angle=0]{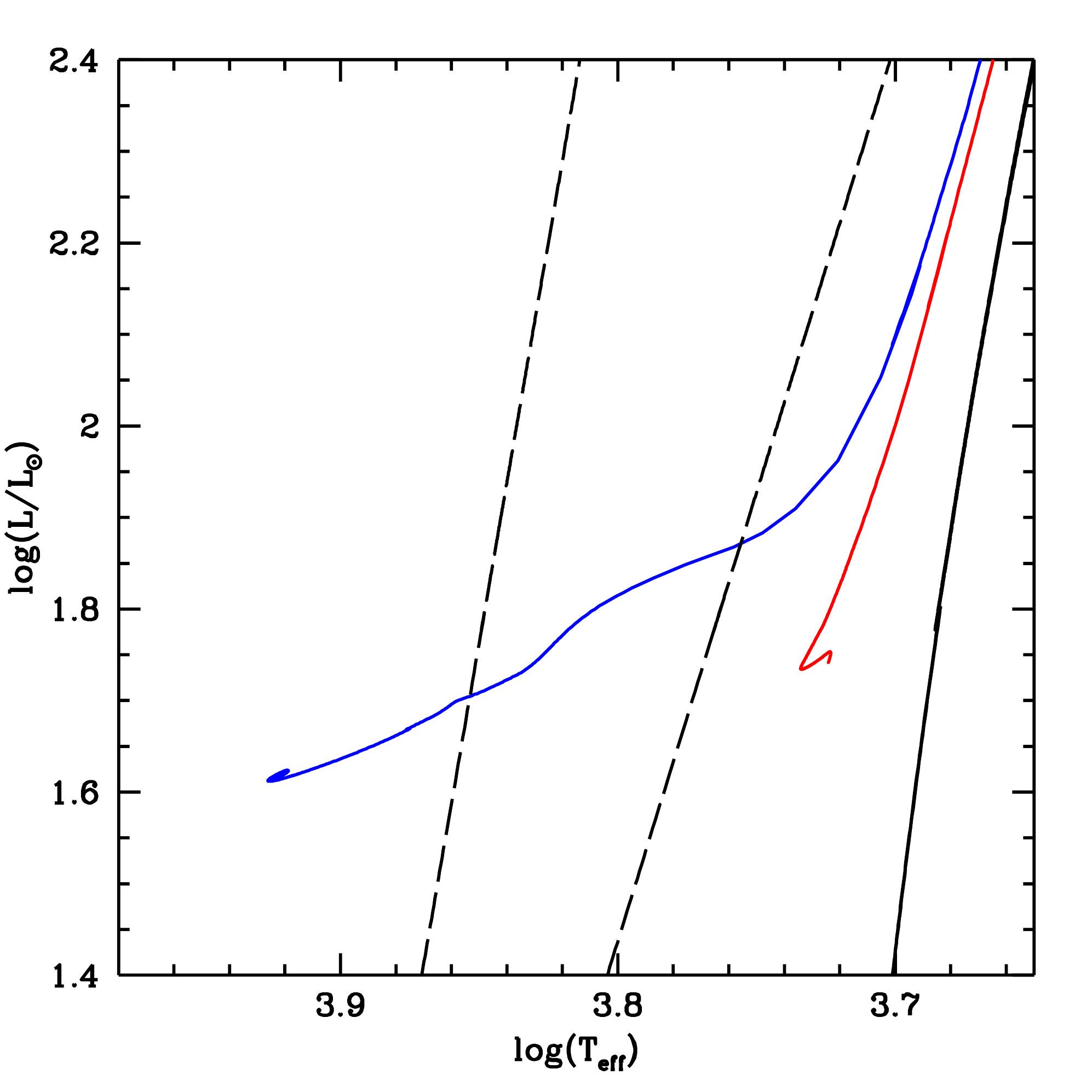}
\caption{A Hertzsprung-Russel diagram showing the zero age horizontal branch position of two stars that have experienced Reimers mass loss on the RGB with $\eta=0.2$ (red line) and $\eta=0.4$ (blue line). The dashed black lines outline the instability strip and the solid black line denotes the location of the RGB for both stars.\label{fig:ZAHB}}
\end{center}
\end{figure}

\begin{figure}
\begin{center}
\includegraphics[width=0.5\textwidth,angle=0]{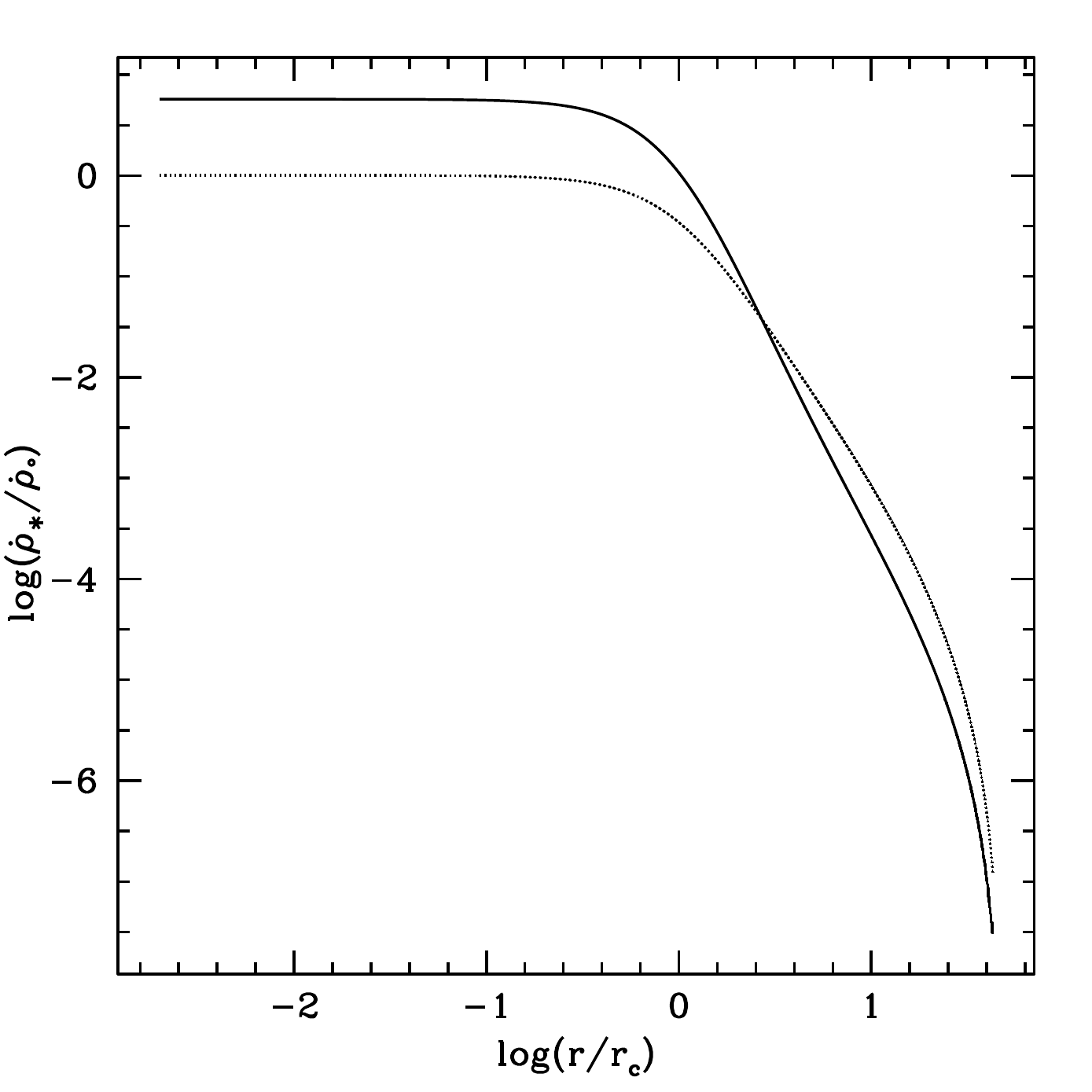}
\caption{The radial profile of the overall local mass-loss rate for the multi-mass (solid line) and King (dotted line) GC models. The mass-loss rates are normalised to the King model's central mass-loss rate.\label{fig:mass-lossProfile}}
\end{center}
\end{figure}

In order to construct a multi-mass GC model, one requires a stellar population created from an initial mass function (IMF). We use the Kroupa IMF \citep{Kroupa02}: Equation (\ref{equ:IMF})
\begin{equation}\label{equ:IMF}
\begin{split}
\frac{dn}{dm}&\,=\,\begin{cases}
&\xi0.019\left(\frac{m}{\Msun}\right)^{-2.3}\;\;    \text{for $m\geq0.5\Msun$}\;,\\
&\xi0.038\left(\frac{m}{\Msun}\right)^{-1.3}\;\;   \text{for $m < 0.5\Msun$}\;,
\end{cases}\\
\text{where}&\\
\xi&\,=\,\begin{cases}
&30.55\times10^5\;\; \text{for the $10^5\Msun$ GC}\;,\\
&30.55\times10^6\;\; \text{for the $10^6\Msun$ GC}\;,
\end{cases}
\end{split}
\end{equation}
to obtain the number of stars and the total stellar mass in each population bin. From this, we obtain a median stellar mass for that bin. We construct a multi-mass King model following the description in \citet{DaCostaFreeman76} and \citet{CapuzzoDolcetta_et_al05}. In summary, this method first constructs a King model GC and fills the resultant gravitational potential with a number of so-called sub-King models, each representing a different mass bin in the stellar population. These sub-King models are related to each other by assuming an equipartition of kinetic energy between them:
\begin{equation}
m_1\sigma_{1}^2\,=\,m_{\rm i}\sigma_{\rm i}^2\;\;\;{\text{where}}\;\mathrm{i}\geq2\;.
\label{equ:equipartition}
\end{equation}
We arbitrarily split our stellar population into ten mass bins where the heaviest two represent the AGB and RGB stellar populations (bins $1$ and $2$, respectively). The rest are split among the sub-giant branch and main sequence stars. A comparison of the King and multi-mass GC models is shown in Fig. \ref{fig:MMGCmodel}.
Using a theoretical isochrone from the BaSTI database \citep{Pietrinferni_et_al04}, we normalise the IMF such that the resulting stellar population represents a currently evolving GC. In order to remain consistent with the gas physics used in VF77, we employ an isochrone with scaled solar metallicity ($Z=0.002$) at an assumed age of $13\Gyr$. This yields a GC where stars more massive than the AGB population ($\sim0.83\Msun$) no longer exist within the cluster (except for stellar remnants such as neutron stars and white dwarfs). We also use this isochrone to determine the stellar properties of this population.\\
\indent As stated before, the purpose of using a multi-mass King model is to better represent the mass loss that occurs within GCs. Each of the RGB and AGB stellar populations therefore have their own specific mass-loss rate, $\alpha_{\small \mathrm{RGB}}$ and $\alpha_{\small \mathrm{AGB}}$. To achieve this, we appropriately sample the isochrone and obtain random stellar properties for each star in our RGB and AGB population bins (Equation (\ref{equ:IMF})). These are then inserted into the relevant empirical mass-loss formula \citep{Reimers75,VassiliadisWood93} to obtain individual mass-loss rates. It is important to note that the Reimers mass-loss formula has a free parameter, $\eta$, which typically adopts the values $0.2$ or $0.4$, where the larger value leads to a higher rate of mass loss. The RGB and AGB specific mass-loss rates are determined by dividing each population's combined mass-loss rate by its total stellar mass. The equivalent global $\alpha$ value is determined by dividing the overall mass-loss rate (i.e. from both the AGB and RGB populations) by the GC mass. It is interesting to ask how such an $\alpha$ value compares to that used by VF77. With $\eta=0.2$, $\alpha=4\times10^{-20}\ps$, an order of magnitude less than that adopted by VF77 and coincidentally matching the observationally determined value in \citet{TaylerWood75}. $\eta=0.4$ yields a global $\alpha$ of roughly $1\times10^{-19}\ps$, one fourth the value adopted by VF77. Adopting $\eta=0.2$ leads to $0.1\Msun$ of gas lost during the RGB phase and a total cluster mass-loss rate of $1.3\times10^{-6}\Msun\pyr$ (for a $10^6\Msun$ GC), of which $58\%$ is from the RGB population and $42\%$ from the AGB. Taking $\eta=0.4$ leads to $0.2\Msun$ lost on the RGB and a total mass-loss rate of $3.5\times10^{-6}\Msun\pyr$, $85\%$ of which is due to the RGB stars and $15\%$ is from the AGB. \citet{McDonald_et_al09} determine a total cluster mass-loss rate of $1.2^{+0.6}_{-0.5}\times10^{-5}\Msun\pyr$ for $\omega$ Centauri and suggest that $0.2-0.25\Msun$ of gas is lost on the RGB. Using a cluster mass  of $3.1\pm0.3\times10^6\Msun$ \citep{Miocchi10}, this corresponds to $\alpha_{\rm \omega\;Cen}=1.2^{+0.8}_{-0.6}\times10^{-19}\ps$, which favours higher mass-loss rates, but is consistent with our derived $\alpha$ values.\\
\indent The resultant zero age horizontal branch (ZAHB) position of a star having undergone our two different mass-loss rates is shown in Fig. \ref{fig:ZAHB}.
The diagram shows how $\eta=0.4$ produces a blue ZAHB star with very little of its envelope remaining. This implies that the value of $\alpha$ adopted by VF77 is perhaps too extreme, resulting in a ZAHB star that will have lost all of its envelope. An $\alpha$ derived from observations is cluster dependent and requires an adopted mass-to-light ratio; this results in an unrelated range of possible values that are hard to apply generally \citep[c.f. the values in][]{Knapp_et_al73,TaylerWood75}. We prefer the approach adopted here; it is better suited for modelling mass loss in GCs and investigating the subsequent ICM evolution in a consistent and methodical way.\\
\indent With consistently determined specific mass-loss rates, we can test how modifying the distribution of mass loss (i.e. the difference between using a King model and a multi-mass GC model) affects the GC ICM evolution. For the multi-mass model, the RGB and AGB stars are more centrally concentrated, resulting in a larger proportion of gas being injected deeper within the potential well of the cluster, as shown in Fig. \ref{fig:mass-lossProfile}. Therefore, for a given ram pressure, we expect less gas to be removed from the GC potential compared to the King-model simulation. The result, is a poorer agreement between our simulations and observations! As the first-ever investigation describing mass loss both in terms of empirical laws and distribution within the cluster, it is essential to ascertain how much the ICM evolution depends upon the GC model.
The subsequent evolution of these two cluster models moving through a Galactic halo medium is discussed in Section \ref{subsec:MMGCRslts}.

\subsection{RGB and AGB mass-loss simulations}
\label{subsec:MMGCRslts}

\begin{figure}
\begin{center}
\includegraphics[width=0.5\textwidth,angle=0]{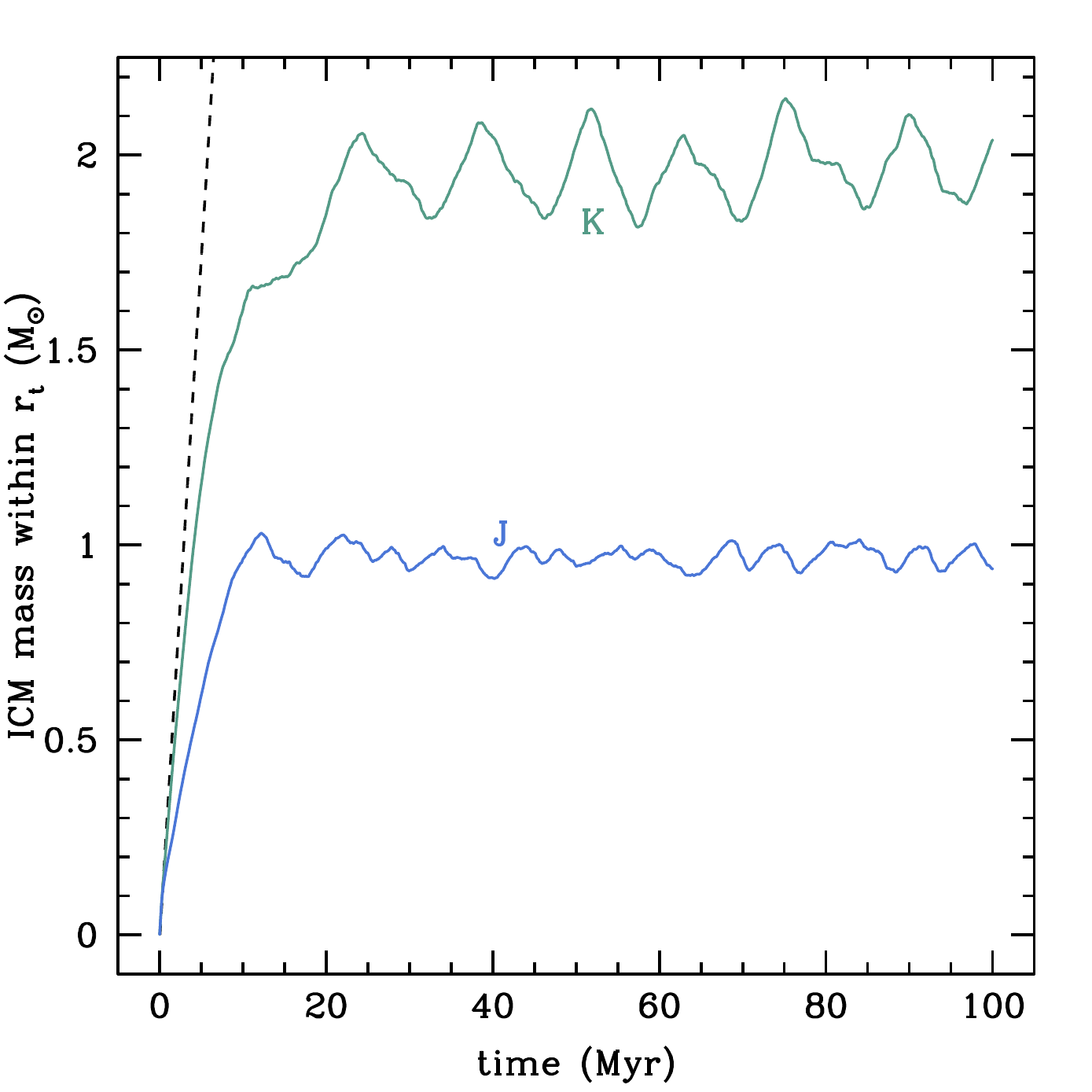}
\caption{The ICM content over time for two $10^5\Msun$ GCs moving through a Galactic halo medium, one modelled with a King model (\textbf{J}) and the other with a multi-mass model (\textbf{K}). The representative stellar populations have the same total mass-loss rate (equivalent to a specific mass-loss value of $\alpha=1\times10^{-19}\ps$). The black dotted line is the injected gas mass against time.\label{fig:JK_compare}}
\end{center}
\end{figure}

\begin{figure}
\begin{center}
\includegraphics[width=0.5\textwidth,angle=0]{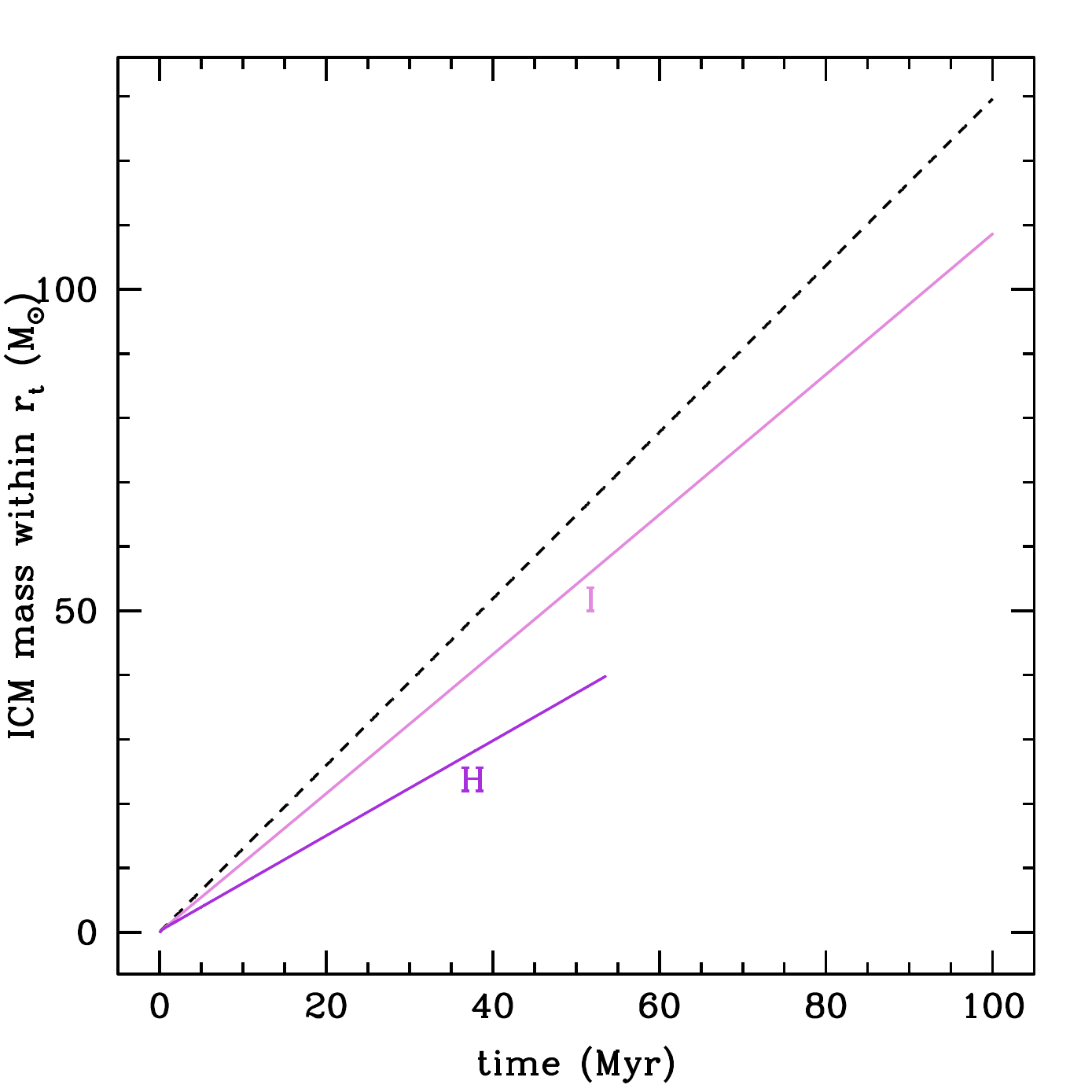}
\caption{The ICM content over time for two $10^6\Msun$ GCs moving through a Galactic halo medium, one modelled with a King model (\textbf{H}) and the other with a multi-mass model (\textbf{I}). The representative stellar populations have the same total mass-loss rate (equivalent to a specific mass-loss value of $\alpha=4\times10^{-20}\ps$). The black dotted line is the injected gas mass against time.\label{fig:HI_compare}}
\end{center}
\end{figure}

\begin{figure}
\begin{center}
\includegraphics[width=0.5\textwidth,angle=0]{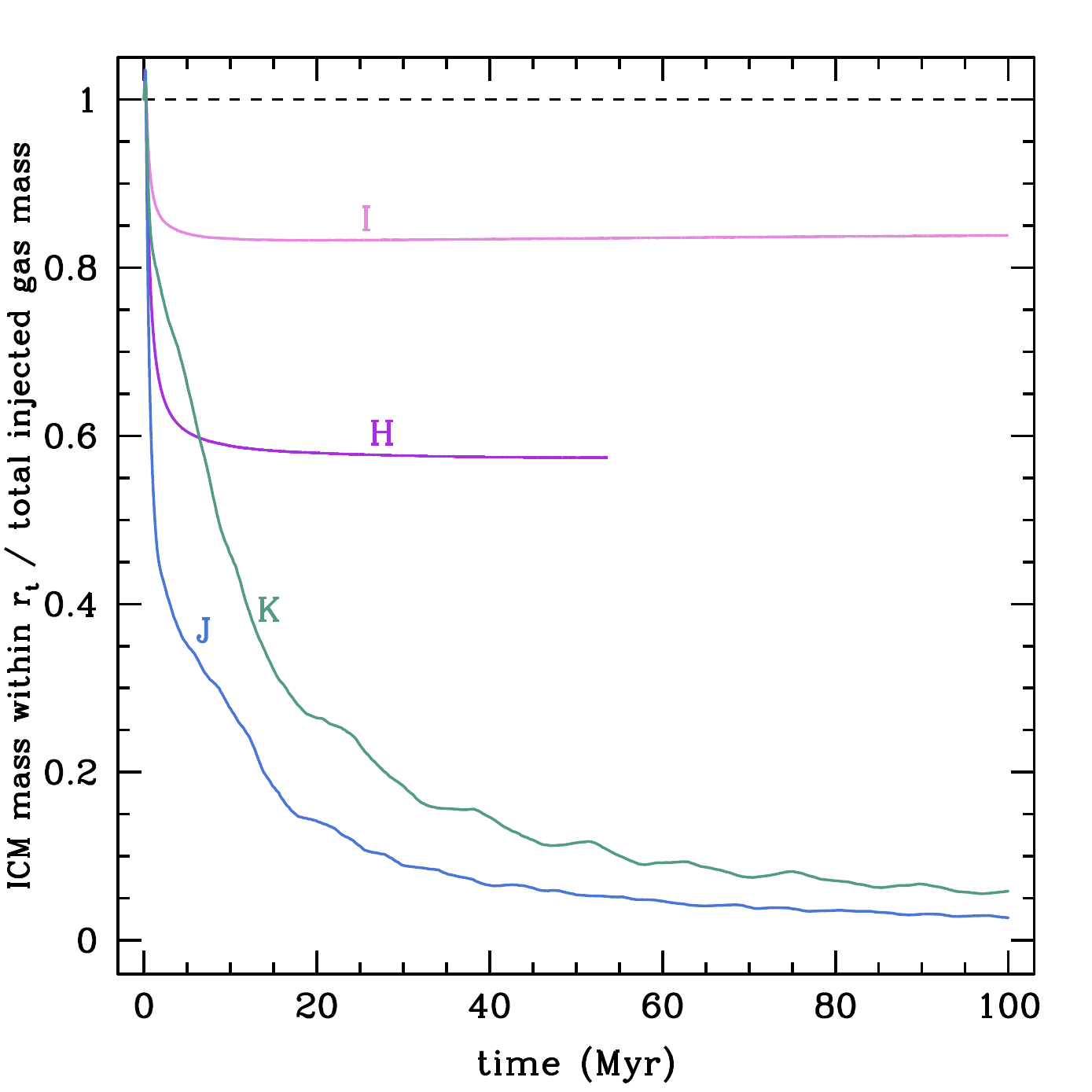}
\caption{The ICM mass within different GC simulations as a fraction of the cumulative injected gas mass against time.\label{fig:HIJK_relative}}
\end{center}
\end{figure}

\begin{figure*}
\begin{center}
\includegraphics[width=0.7\textwidth,angle=0]{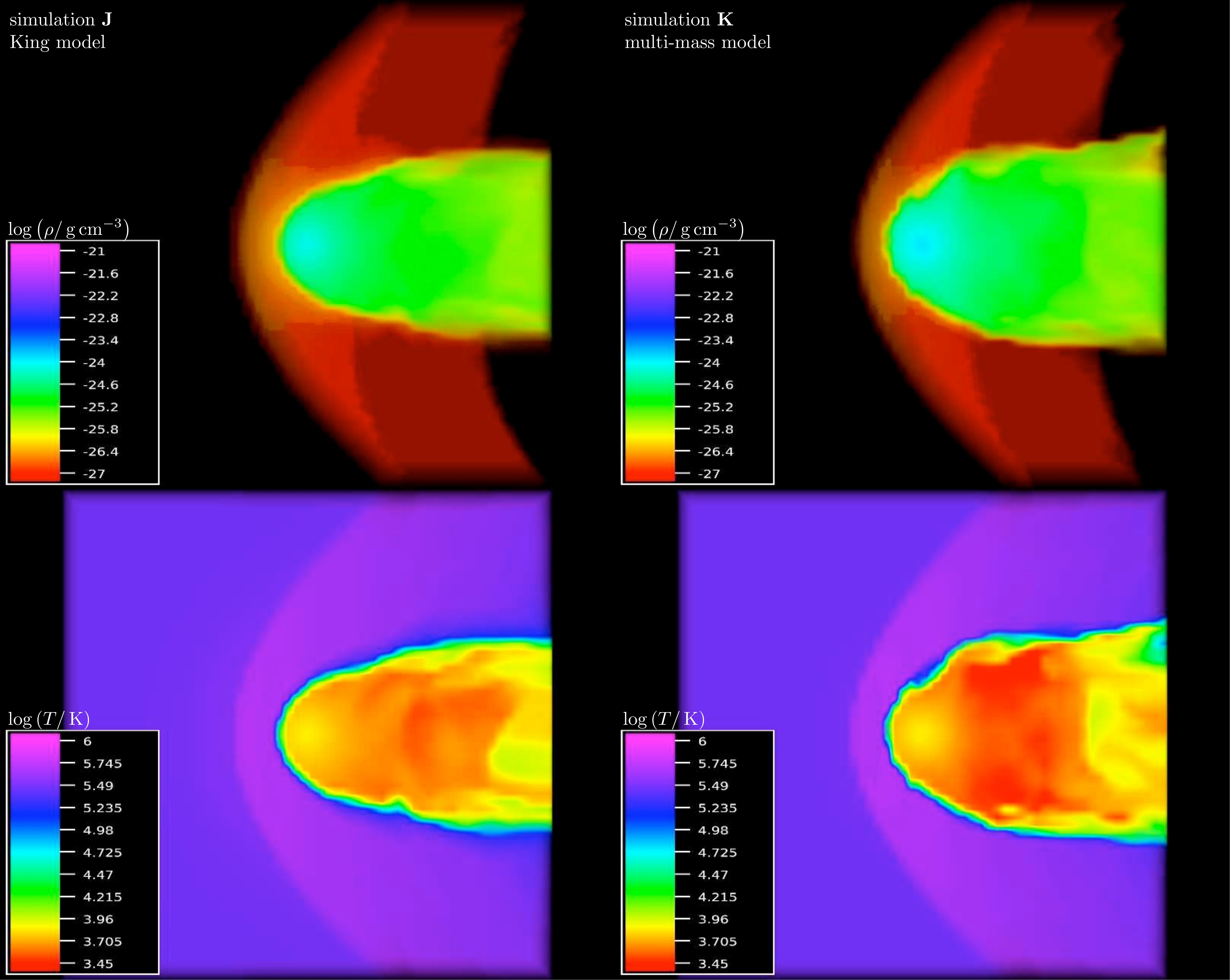}
\caption{A cutaway of 3D rendered images showing $\log\left(\rho/\gcc\right)$ (upper panels) and  $\log\left(T/\K\right)$ (lower panels) of the ICM gas at $10\Myr$ for a $10^5\Msun$ GC represented by a King model (simulation \textbf{J}; left panels) and multi-mass model (simulation \textbf{K}; right panels). These data are taken from the $3^{\rm rd}$ grid that has a side length of $24.5\pc$.\label{fig:JK_colourmap}}
\end{center}
\end{figure*}

\begin{figure*}
\begin{center}
\includegraphics[width=0.7\textwidth,angle=0]{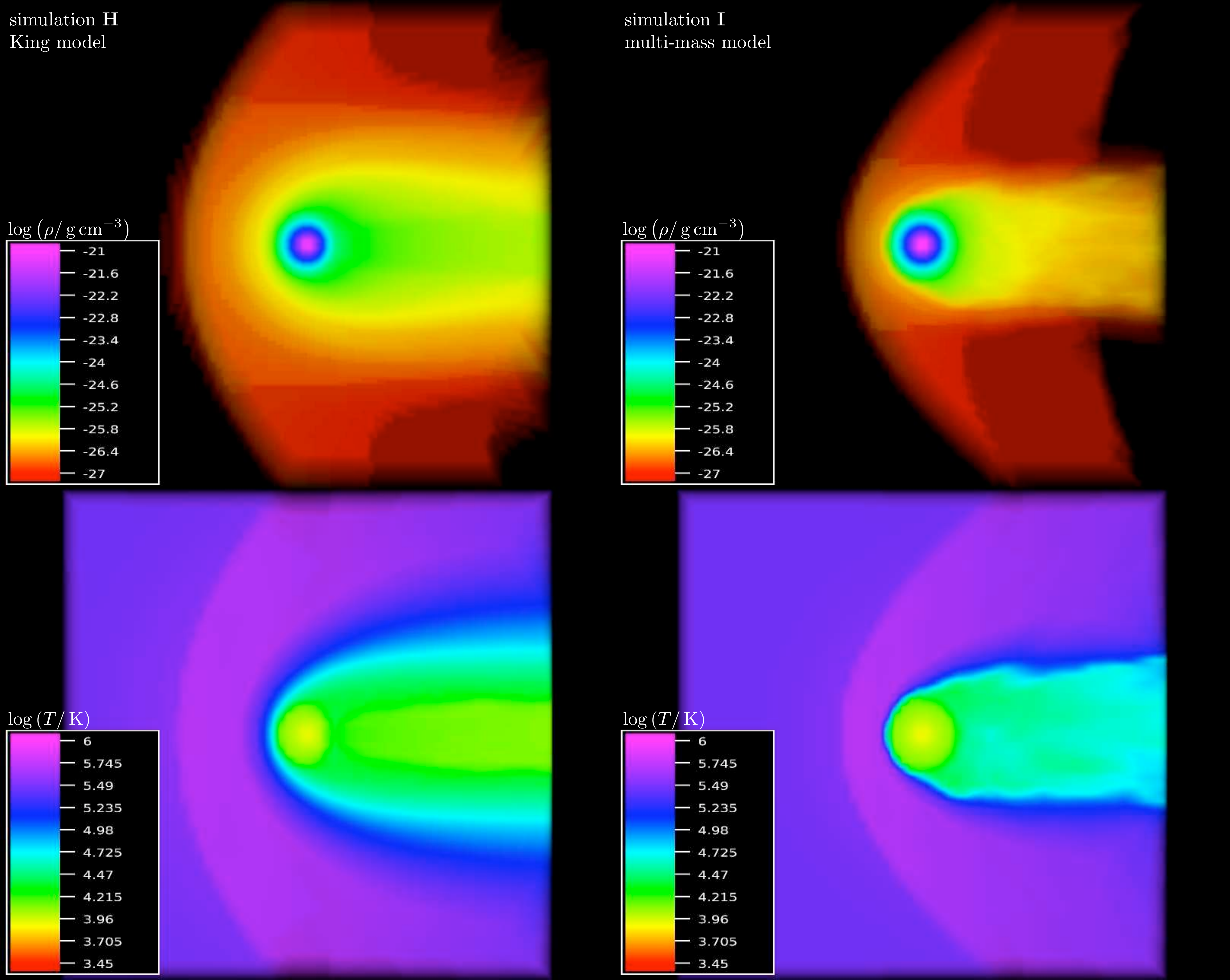}
\caption{A cutaway of 3D rendered images showing $\log\left(\rho/\gcc\right)$ (upper panels) and  $\log\left(T/\K\right)$ (lower panels) of the ICM gas at $10\Myr$ for a $10^6\Msun$ GC represented by a King model (simulation \textbf{H}; left panels) and multi-mass model (simulation \textbf{I}; right panels). These data are taken from the $3^{\rm rd}$ grid that has a side length of $24.5\pc$.\label{fig:HI_colourmap}}
\end{center}
\end{figure*}

We run two high-resolution simulations each for a $10^5\Msun$ ($64^3$ with seven nested grids) and $10^6\Msun$ ($64^3$ with six nested grids) GC in order to compare the difference between using a King model GC and a multi-mass model GC (i.e. four simulations in total); these are summarised in Table \ref{tab:MMsimulations}. 
\begin{table}\centering
  \setlength\extrarowheight{2pt}
  \caption[]{Summary of the simulations that compare the ICM evolution using a King model GC and a multi-mass GC.}
  \label{tab:MMsimulations}
  \begin{tabular}{cccc}
    \\
    \hline
    Simulation
    & $M_{\mathrm{GC}}$ $\left(\Msun\right)$ & GC model
    & $\alpha$ $\left(\ps\right)$ \\
    \hhline{====}
    \textbf{H} & $10^6$ & King   & $4.01\times10^{-20}$\\
    \textbf{I} & $10^6$ & multi-mass & ---------\\
    \textbf{J} & $10^5$ & King & $1.08\times10^{-19}$\\
    \textbf{K} & $10^5$ & multi-mass & ---------\\
    \hline
    \\
  \end{tabular}
\end{table}
The method used to produce the initial King model that is involved in constructing our multi-mass models, is slightly different to the method employed for the King models described in Sections \ref{sec:numerical_methods} and \ref{sec:halo_motion}. Therefore, the King models used in this section and previous sections are comparable, but not identical. The parameters that describe the new King models are provided in Table \ref{tab:MM_king_GC}.
\begin{table}\centering
  \setlength\extrarowheight{2pt}
  \caption[Properties of the new globular cluster models]{Properties of the two King models created for the purpose of constructing the multi-mass GC models.}
  \label{tab:MM_king_GC}
  \begin{tabular}{cccc}
    \\
    \hline
    $M_{\mathrm{GC}}$ $\left(\Msun\right)$ & $\rho_\circ$ $\left(\gcc\right)$ & $\rc$ $\left(\pc\right)$
    & $\rt$ $\left(\pc\right)$\\
    \hhline{====}
    $1.03\times10^5$ & $1.653\times10^{-18}$ & $0.51$ & $24.5$\\
    $1.03\times10^6$ & $1.653\times10^{-17}$ & $0.51$ & $24.5$\\
    \hline
    \\
  \end{tabular}
\end{table}
We run a new set of King-model GC simulations in order to maintain a consistent comparison with our multi-mass GC simulations. The initial conditions for all simulations are the same as in models \textbf{D} and \textbf{F} ($10^6\Msun$ and $10^5\Msun$ GCs respectively). That is, a Halo temperature of $10^{5.5}\K$, a Halo density of $10^{-27}\gcc$ and a GC velocity of $200\kms$. Since it is relatively easy to strip gas from the $10^5\Msun$ GCs, we use the Reimers mass-loss law with $\eta=0.4$ (corresponding to $\alpha=1\times10^{-19}\ps$). For the $10^6\Msun$ GC, we use the lower extreme of $\eta=0.2$ ($\alpha=4\times10^{-20}\ps$) in order to maximise gas stripping. In the following discussion, we compare how the ICM evolution changes between the King GC model and the multi-mass description of the cluster. We first compare the $10^5\Msun$ GC simulations (\textbf{J} and \textbf{K}, Fig. \ref{fig:JK_compare}) and then discuss the results from the $10^6\Msun$ simulations (\textbf{H} and \textbf{I}, Fig. \ref{fig:HI_compare}). Lastly, we compare the relative evolution of the ICM mass for all simulations (Fig. \ref{fig:HIJK_relative}).\\
\indent In Section \ref{subsec:HaloRslts}, the long-term build-up of ICM gas in $10^5\Msun$ clusters was prevented, due to the ram pressure experienced as the GC moved through the ambient medium. Fig. \ref{fig:JK_compare} shows this is also the case for the multi-mass cluster. Compared to the King model (\textbf{J}), however, more gas is retained by the GC potential before the ICM content plateau (at $\sim24\Myr)$. The plateau mass of gas within the cluster for the rest of the simulation is roughly $1.96\Msun$, fluctuating by up to $9\%$. This fluctuation is due to the central reservoir of gas increasing in size until Kelvin-Helmholtz (KH) instabilities develop on the surface tangential to the flow of Halo gas around the ICM gas. These KH instabilities result in an episode of enhanced gas stripping, which reduces the size of the core. This cycle of ICM growth and shedding occurs on timescales of $13\pm1.5\Myr$. The King model reaches a plateau of $0.97\Msun$ in half the time and can vary by up to $6\%$. In this steady-state condition, the multi-mass simulation has just over twice as much gas compared to the King-model simulation. The corresponding hydrogen content within $\rt$ for these two model clusters are $0.68\Msun$ and $1.37\Msun$ for \textbf{J} and \textbf{K}, respectively, which are above observational limits. However, within a core region of $2\pc$, there is as little as $0.15\Msun$ and $0.29\Msun$ of hydrogen gas present, about the same order as the most stringent observational upper limits, but much less than most upper limits. Again, most of the gas within $\rt$, resides within the tail of stripped material, where photometric detection will be line-of-sight dependent.
\\
\indent For the $10^6\Msun$ model simulations, the multi-mass GC also retains more gas than the King-model simulation. Fig. \ref{fig:HI_compare} shows that simulation \textbf{H} develops similarly to simulation \textbf{E} (Fig. \ref{fig:BE_compare}), the rate of gas stripping due to ram pressure being stable over tens of $\Myr$. This results in the ICM mass increasing linearly at about $57\%$ the rate of injection of stellar material. The multi-mass model (simulation \textbf{I}) experiences a much lower rate of gas stripping, with the ICM increasing at about $84\%$ of the rate of stellar mass loss ($47\%$ more than \textbf{H}). Since the ICM content increases continually with time, we again choose to assess the hydrogen content at $10\Myr$. Model \textbf{H} contains $5.0\Msun$ within $\rt$, of which $4.6\Msun$ resides within the central $2\pc$.  In comparison, simulation \textbf{I} contains $7.7\Msun$ of hydrogen, $7.6\Msun$ residing within $2\pc$. Therefore, we expect to observe a central reservoir of gas for massive clusters moving through the tenuous upper Halo.
\\
\indent The relative ICM content, compared to the injected mass, from all simulations is shown in Fig. \ref{fig:HIJK_relative}. It highlights well, the significant influence that cluster mass has on the mass of retained ICM. From these simulations, it appears that removing gas from the gravitational potential of massive GCs may be a bigger challenge than previous studies have inferred. The $\sim50\%$ difference ($100\%$ in the case of $10^5\Msun$ clusters) in ICM content generated by altering the distribution of stellar mass loss re-enforces the conclusion of Section \ref{sec:halo_motion}: it is important to appropriately model the stellar population and associated mass loss.\\
\indent We present a snapshot at $10\Myr$ of the gas density and temperature for the $10^5\Msun$ and $10^6\Msun$ GCs in Fig. \ref{fig:JK_colourmap} and Fig. \ref{fig:HI_colourmap}, respectively. These figures show the morphological differences that emerge from using the King model (left panel of Fig. \ref{fig:JK_colourmap} and Fig. \ref{fig:HI_colourmap}) and the multi-mass one (right panel). The multi-mass GC simulations, show a larger volume of denser material in the core compared to the King-model simulation. The $10^5\Msun$ simulations show a cool tail of stripped gas (which contains neutral hydrogen) with no evidence of a distinct, bound core of gas in the central regions. In contrast, the $10^6\Msun$ simulations show a central mass of gas that is thermally distinct from the tail of stripped material, indicating that this gas is bound and cools over time. The hydrogen in this dense central reservoir is atomic.

\section{Discrete stellar population cluster model}
\label{sec:discrete_stars}

Section \ref{subsec:MMGCRslts} highlighted the influence the GC model has on simulating ICM evolution. The King and multi-mass GC models both represent the stellar population as a continuous stellar density distribution. This approximation is upheld in the limit of a large number of stars per unit volume. This is indeed the case (although is less valid at the cluster outskirts) for the King model, which has a homogeneous stellar population. For a multi-mass model, this approximation remains valid for the numerous ($10^5-10^6$) main-sequence stars. However, the approximation of a continuous stellar distribution breaks down for the RGB and AGB populations, since they account for less than a percent of the total population. For our $10^5\Msun$ GC, the IMF produces $814$ RGB stars and just $2$ AGB stars. The $10^6\Msun$ GC contains $8140$ and $20$ of each, respectively. Modelling these populations as individual stars orbiting within the GC potential will alter the distribution of gas injection from a monolithic, symmetric one  to a time-dependent distribution of asymmetrically dispersed points. In such a distribution, the ICM injection is not centred on the deepest parts of the gravitational potential, allowing a larger fraction to be liberated by ram pressure. A distribution of independent sources of material, increases the effective surface area upon which ram pressure will act to lift gas out of the potential. Furthermore, an asymmetric ICM morphology will provide more seed perturbations from which KH instabilities can grow, leading to the mixing and fragmentation of the ICM. The importance of KH instabilities on the fragmentation and break-up of a cloud moving through an ambient medium, has already been demonstrated in \citet{Agertz_et_al07}. Although our ICM gas is continuously replenished, we hypothesise that these effects will have a significant impact on ICM evolution, leading to a greater level of gas stripping. In the following sections, we present our discrete GC model and subsequent simulations.

\subsection{Rationale, setup and initial conditions}
\label{subsec:discrete_setup}

For the purpose of numerical efficiency, we make a few simplifying approximations in modelling our discrete GCs. We choose to represent our RGB and AGB populations as point sources of mass loss, moving as test particles within the smooth GC potential. Since we are only concerned with the stellar winds of these two populations, modelling the other stars as discrete objects is unnecessary. Secondly, by neglecting full N-body calculations, we avoid consuming CPU time on resolving unrelated binary interactions. Furthermore, ignoring star-star interactions, isolates the influence of the discretisation and stellar motion (i.e. we don't include the effect of a noisy, time-varying potential).\\
\indent We build our model by giving each star, from each population, an initial position and velocity within our simulation volume. We use the multi-mass GC model in spherical polar coordinates as the framework for obtaining these initial values via Monte-Carlo techniques. The angular positions of our stars are distributed evenly, whereas the radial position is weighted according to the sub-King model for each population. At each radial position, the magnitude of the stellar velocity is taken to be the local root mean squared velocity and the vector direction is chosen at random. Each star is given its own individual empirical mass-loss rate (the same as that used in calculating the specific mass-loss rates in Section \ref{sec:MMGC}).\\
\indent The procedure used to employ the discrete-model GCs into the simulations is slightly modified compared to that described in Section \ref{sec:numerical_methods}. For the injection of stellar material, each cell no longer holds a stellar density value. Instead, we track the position of each star, relative to the grid and add gas directly into the cell it resides in. Cells that contain no stars do not have material injected into them. The local ICM mass, momentum and energy properties are updated accordingly (Section  \ref{sec:numerical_methods}). The local momentum and energy conservation Equations, (\ref{equ:Kmomentum_inj}) and (\ref{equ:Kenergy_inj}) respectively, must be generalised slightly to include the extra terms that vanish in the limit of an isotropic, continuous stellar distribution. The local momentum is conserved according to
\begin{equation}
\rho_{\mathrm{n+1}}\bm{v}_{\mathrm{n+1}}\,=\,\rho_{\mathrm{inj}}\bm{v}_{\mathrm{\ast}}+\rho_{\mathrm{n}}\bm{v}_{\mathrm{n}}\;,
\label{equ:MMmomentum_inj}
\end{equation}
and the energy injected by the stellar wind, $U_{\alpha}$, is modified such that
\begin{equation}
U_{\alpha}\,=\,\beta+\frac{1}{2}v_{\ast}^{2}-\bm{v}_{\ast}.\bm{v}_{\mathrm{n+1}}+\frac{1}{2}v_{\mathrm{n+1}}^{2}\;.
\label{equ:MMUalpha_inj}
\end{equation}
After injecting mass onto the grid, the stars are moved (with the hydrodynamic timestep) under the influence of gravity, using the Euler method and a predictor-corrector scheme. The rest of the hydrodynamic calculations follow the sequence described in Section \ref{sec:numerical_methods}.

\subsection{Evolution of the ICM within a discrete stellar environment}
\label{subsec:DiscreteResults}

\begin{figure}
\begin{center}
\includegraphics[width=0.5\textwidth,angle=0]{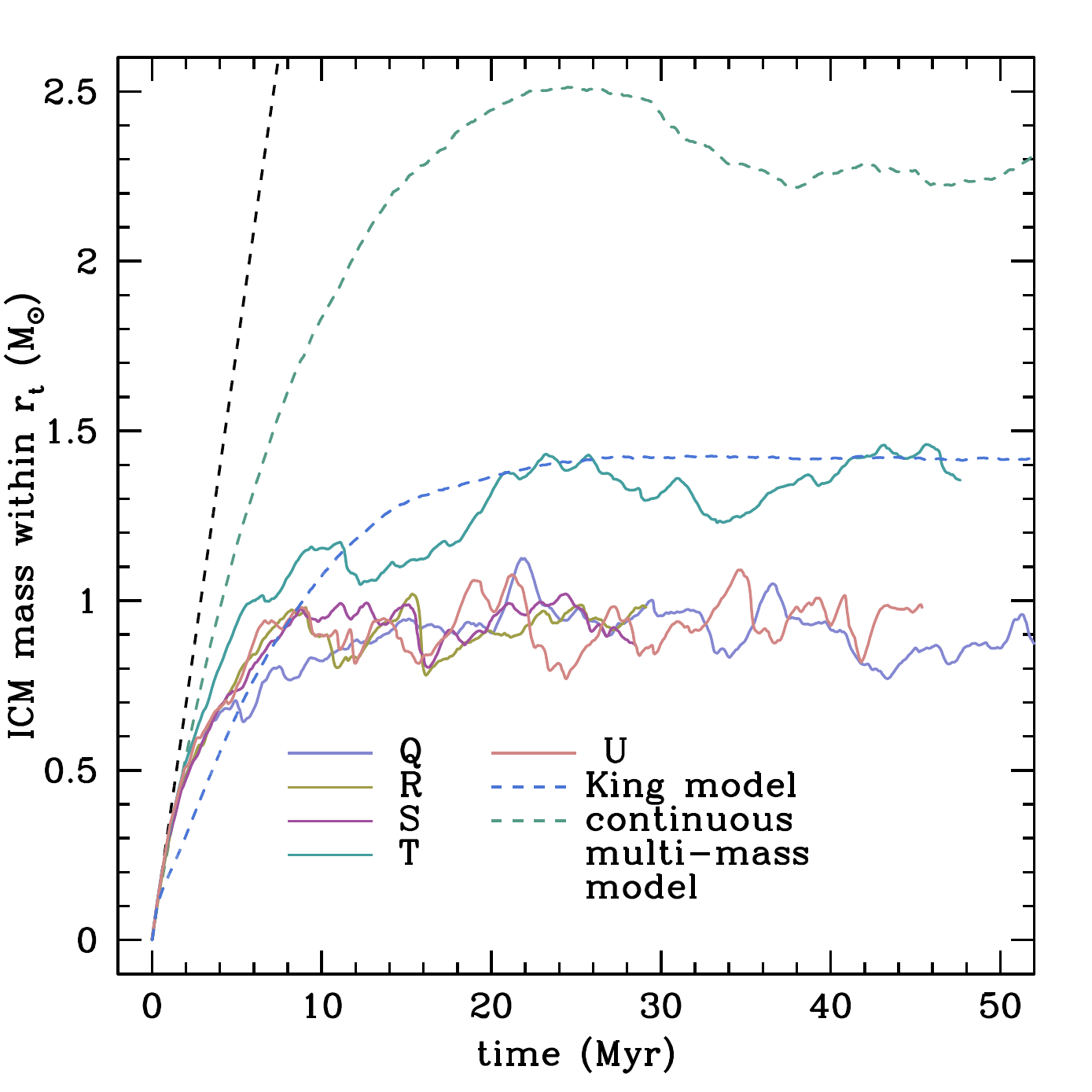}
\caption{The ICM content over time for seven $10^5\Msun$ GCs moving through a Galactic halo medium. Five are represented by a discrete multi-mass model with different initial stellar positions and velocities (simulations \textbf{Q} through \textbf{U}; solid coloured lines), one is represented by a King model and another is represented by a continuous multi-mass model. The representative stellar populations in each GC model have the same total mass-loss rate (equivalent to a specific mass-loss value of $\alpha=1\times10^{-19}\ps$). The black dotted line is the injected gas mass against time.\label{fig:QRSTU_compare}}
\end{center}
\end{figure}

\begin{figure*}
\begin{center}
\includegraphics[width=0.7\textwidth,angle=0]{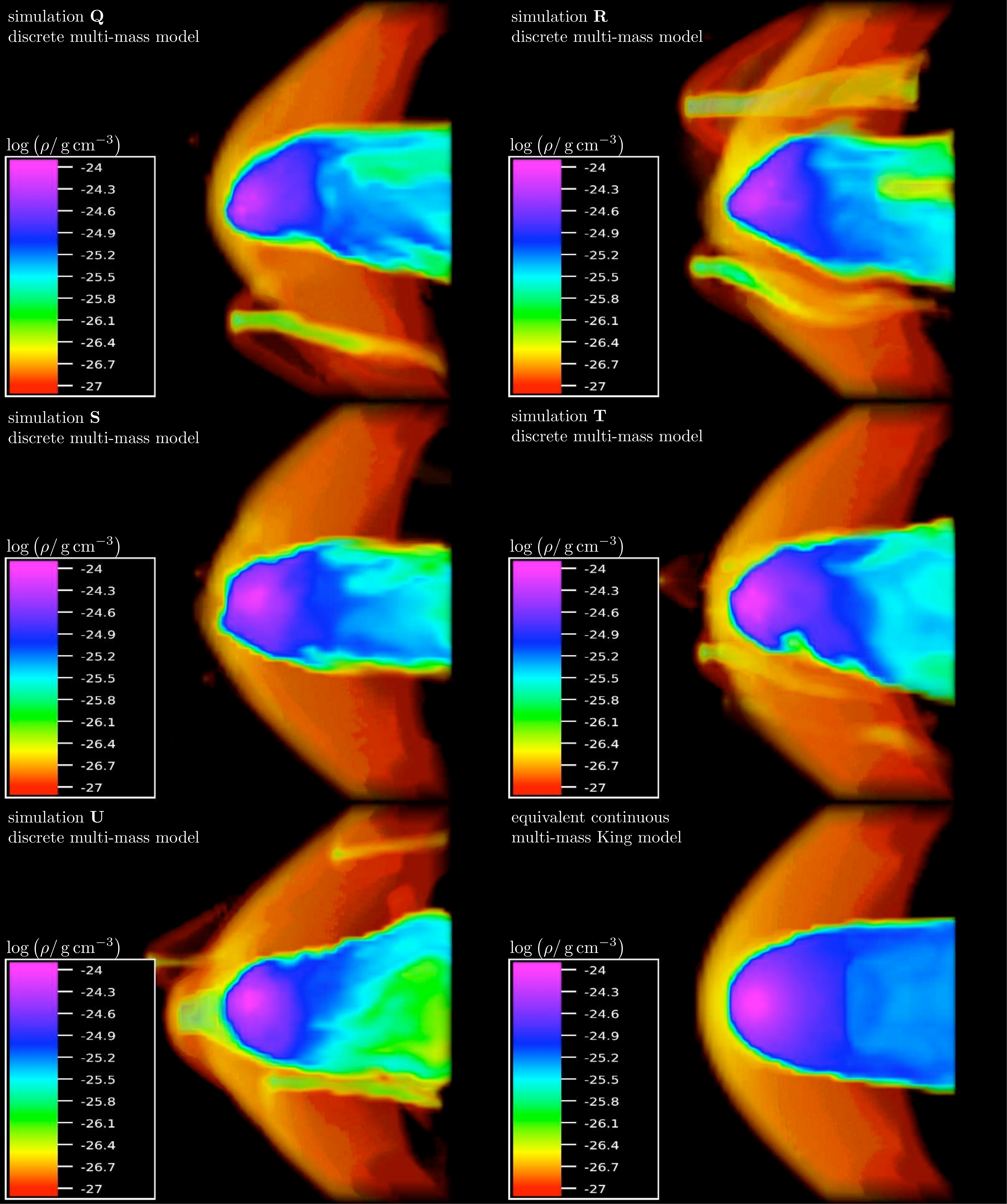}
\caption{A cutaway of 3D rendered images showing $\log\left(\rho/\gcc\right)$ of the ICM gas at $10\Myr$ for a $10^5\Msun$ GC, represented by discrete multi-mass GC models that have different initial stellar positions and velocities (simulations \textbf{Q}, \textbf{R}, \textbf{S}, \textbf{T}, and \textbf{U}). The bottom-right panel is from an equivalent continuous multi-mass simulation. These data are taken from the $3^{\rm rd}$ grid which has a length of $24.5\pc$.\label{fig:QRSTU_dnsmap}}
\end{center}
\end{figure*}

\begin{figure*}
\begin{center}
\includegraphics[width=0.7\textwidth,angle=0]{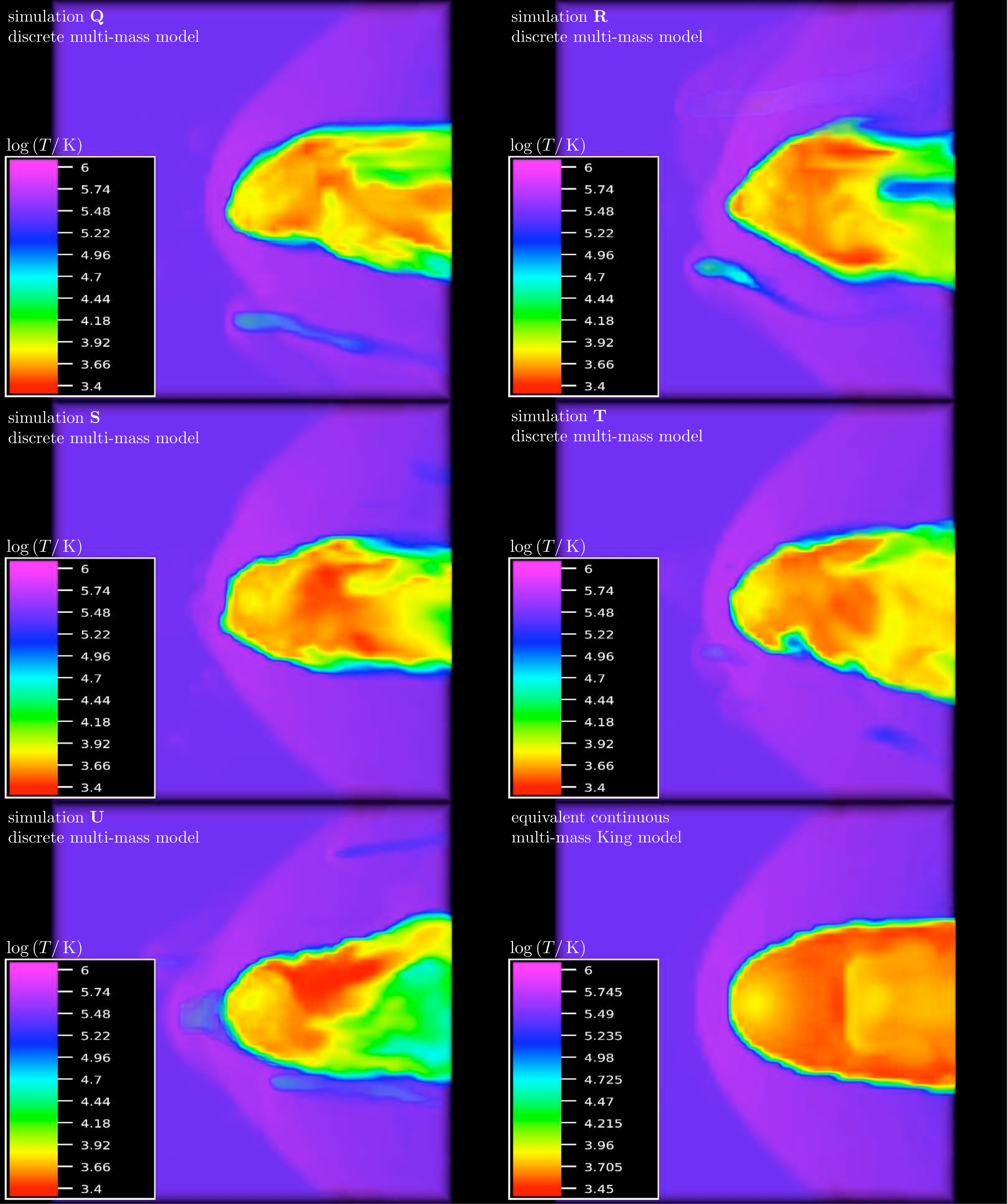}
\caption{A cutaway of 3D rendered images showing $\log\left(T/\K\right)$ of the ICM gas at $10\Myr$ for a $10^5\Msun$ GC, represented by discrete multi-mass GC models that have different initial stellar positions and velocities (simulations \textbf{Q}, \textbf{R}, \textbf{S}, \textbf{T}, and \textbf{U}). The bottom-right panel is from an equivalent continuous multi-mass simulation. These data are taken from the $3^{\rm rd}$ grid which has a length of $24.5\pc$.\label{fig:QRSTU_tmpmap}}
\end{center}
\end{figure*}

For each GC mass, we run a total of five low-resolution simulations using three nested grids, \textbf{L} to \textbf{P} for the $10^6\Msun$ GC and \textbf{Q} to \textbf{U} for the $10^5\Msun$ GC. For all simulations, the initial conditions for the gas on the grid are $\rho_{\rm H}=10^{-27}\gcc$, $T_{\mathrm{H}}=10^{5.5}\K$ and $v_{\rm \small GC}=200\kms$. Each simulation has a different set of randomly determined initial positions and velocities for the stars. The Halo initial conditions remain the same as those used in previous sections. We run five simulations in order to see what effect, if any, the choice of initial stellar positions and velocities has on the global ICM evolution. In this section, we present the results from the $10^5\Msun$ simulations, followed by a brief description of those from the $10^6\Msun$ simulations. We then discuss  some of the technical challenges faced when attempting high-resolution simulations and conclude with a summary of the main results from this experiment and discuss the consequences for future investigations.\\
\indent Section \ref{subsec:MMGCRslts} showed that a multi-mass GC model (employing mass-loss laws) results in a higher retention of ICM gas than the King model. The predicted ICM mass from our discrete multi-mass GC simulations are given in Fig. \ref{fig:QRSTU_compare}.
Four of the five discrete GC simulations show a plateau mass of $0.92\Msun$, varying by about $10\%$ due to stellar orbits moving gas injection around the potential well. This is $60\%$ less than a low-resolution continuous multi-mass GC model ($2.28\Msun\pm12.5\%$) and $35\%$ less than an equivalent King-model simulation ($1.43\Msun\pm1\%$). The discrete model that predicts the highest ICM mass (\textbf{T}) is comparable with the King-model simulations at $1.35\Msun\pm10\%$. This simulation retains more gas from the start, implying that some members of its stellar population are in lower eccentricity orbits, spending more time deeper in the potential well compared to other simulations. However, stellar orbits have not been traced throughout the simulations and this hypothesis has not been confirmed.\\
\indent The log density and temperature profiles of these simulations at $10\Myr$ are presented in Fig. \ref{fig:QRSTU_dnsmap} and Fig. \ref{fig:QRSTU_tmpmap}, respectively.
Compared to a low-resolution continuous multi-mass GC simulation (bottom-right panel in Fig. \ref{fig:QRSTU_dnsmap} and Fig. \ref{fig:QRSTU_tmpmap}), the ICM morphology from the discrete simulations show less ICM content due to a narrower cross-sectional profile. The streaming tail of gas shows more structure and a lower density. A few stars are also observed losing mass separately from the main region of ICM gas. A summary of the amount of hydrogen within the central $0.2\pc$ is given in Table \ref{tab:discreteHmass}.
\begin{table}\centering
  \setlength\extrarowheight{2pt}
  \caption{The mass of hydrogen residing within the core of the $10^5\Msun$ GC discrete multi-mass simulations.}
  \label{tab:discreteHmass}
  \begin{tabular}{cc}
    \\
    \hline
    Simulation
    & H mass within $2\pc$ $\left(\Msun\right)$\\
    \hhline{==}
    \textbf{Q} & $0.11$\\
    \textbf{R} & $0.12$\\
    \textbf{S} & $0.15$\\
    \textbf{T} & $0.17$\\
    \textbf{U} & $0.16$\\
    \hline
    \\
  \end{tabular}
\end{table}
These values range from $0.11\Msun$ to $0.17\Msun$ and reside between the $0.1\Msun$ value for the King-model simulations and the $0.29\Msun$ for the continuous multi-mass model.\\
\indent A discrete model for the $10^6\Msun$ GC simulations  produced no change in the overall ICM mass content with time, compared to a low-resolution continuous multi-mass simulation. This is likely because the stellar population is unresolved in these low-resolution simulations. Accordingly, the discrete simulations behave much like the continuous stellar density case. However, it is also possible that, for the massive GC, the number density of RGB and AGB stars is high enough to warrant the use of a continuous stellar density distribution. To determine the effect of discretising the stellar population on the ICM evolution, the stellar population needs to be resolved. This requires higher resolution simulations with a larger number of nested grids so that the individual stars can be separated on the grid. Nevertheless, it is encouraging that these initial $10^5\Msun$ GC simulations indicate that discretising the stellar population has a favourable effect on reducing the level of ICM gas within GCs.\\
\indent  We attempted high-resolution simulations ($64^3$ with six levels of refinement). However, we discovered a numerical problem in our simple discrete GC model. In these simulations extreme temperatures spanning $\sim10^9\K$ to $\sim0.1\K$ are observed. We realised that by injecting the gas into a single cell, we create a numerically unresolved injection process and do not correctly simulate the changes in the gas flow properties as the stellar wind interacts with the Halo gas. This discontinuous peak in density, contained within a solitary cell, is an obstacle that Halo gas cannot flow around properly. This leads to artificially low densities in the star's wake ($\rho\sim10^{-30}\gcc$) as gas leaves these cells due to the relative motion of the Halo. This wake gas is not replenished by the upstream wind material at an appropriate rate since the flow of this material is not established over enough cells. This wake gas not only has a reduced density but is also artificially heated to $\sim10^9\K$, causing numerical errors later on in the simulation such as extremely cold gas. This cold gas has a low internal energy which results in tiny energy injection timesteps and the simulation grinds to a halt. This numerical effect, caused by injecting gas into a single cell, becomes more extreme with increasing resolution and so our initial simulations therefore do not show temperature variations beyond that expected from cooling. Although the gas physics in these low-resolution discrete simulations suffer from numerical artefacts, we firmly believe that this does not change the predicted low levels of ICM gas. The continuous stellar density simulations do not suffer from this numerical issue because the gas is injected over the entire cluster and the flow of this new gas is extended over most of the cells in the simulation volume. Future work will present simulations where the injection of stellar wind material from discrete sources is spread over several cells. The resultant high-resolution simulations aim to confirm our initial findings, as well as determine the effect that discretisation has on the ICM evolution of $10^6\Msun$ GCs.\\
\indent Our initial findings from using a discrete GC model, are presented here as a proof-of-concept result. Breaking up the injection of stellar winds into localised sources, enables gas to escape the cluster potential more easily, as well as encourage the growth and development of instabilities that lead to the fragmentation and dissipation of a replenished intra-cluster medium.

\section{Summary and Discussion}
\label{sec:Summary}

\subsection{Discussion}
\label{subsec:discussion}

Our work highlights the importance of carefully modelling the amount and distribution of mass loss, when simulating the evolution of ICM gas in GCs. Defining the nature of mass loss from population II giants is therefore essential for understanding the evolution of gas within GCs. In this work, we employed the Reimers formula, which is often applied in many fields to describe mass loss from GC RGB stars. However, the Reimers law is calibrated on population I stars and isn't successful at explaining all the observations of GC stellar populations. There are currently only a handful of alternative formulae available that have been calibrated against globular cluster RGB and AGB stars. They give quite different mass-loss rates, some suggesting that these winds are episodic in nature, highlighting the fact that the mechanisms governing mass loss are poorly understood \citep{Origlia_et_al07,Meszaros_et_al09,McDonald_et_al09}. Although on average the same amount of gas must be injected into the ICM, episodic mass-loss may have an encouraging role in removing gas from the cluster potential. Episodic mass-loss will cause variations in the local stellar mass-loss rate due to some stars turning off their mass loss whilst others begin their duty cycle. In those regions where more stars are in an ``off'' phase, a larger portion of ICM gas could be removed by ram pressure per unit time. However, simulation is the only way to determine the influence this sort of mass loss will have on the evolution of the GC ICM.\\
\indent For ``typical'' (i.e. $10^5\Msun$) GCs, we used the higher value for the Reimers stellar mass-loss rate (i.e. $\eta=0.4$) as described in Section \ref{subsec:MMGCRslts}. Even with this higher rate, our predicted ICM content is less than the majority of observational upper limits. Therefore, if we adopt a lower Reimers mass-loss rate ($\eta=0.2$) for stars on the RGB, we would obtain an even lower predicted ICM mass. In contrast, the massive ($10^6\Msun$) GC models used the lower Reimers mass-loss rate and yet still show a disparity between the predicted and observed ICM mass for the massive GC models. We propose that discretising the distribution of mass loss in simulations will improve the gap between theory and observation. However, for $10^6\Msun$ clusters employing higher stellar mass-loss rates, it is likely that other mechanisms, such as those discussed in Section \ref{sec:intro}, will need to be included in order to prevent the build-up of ICM gas within massive GCs.\\
\indent We note that the radial distribution of our stellar population is fixed and does not change with time. Real GCs, however, are dynamical systems in quasi-equilibrium states, where stellar distributions change with time. Events such as the hardening of binary systems and stellar mass loss, result in an expansion of the core; this would make ram-pressure stripping more effective. Since we model mass loss, we must determine whether our adopted static (King and multi-mass) models are justified. To do this, we look at the stellar evolutionary and GC central relaxation timescales, as well as the fraction of stellar mass lost via winds between Disk crossing events. The timescales over which our RGB and AGB stars evolve are of the order $10^8$ years. Similarly, the central relaxation timescales of Galactic GCs show a median of $3.4\times10^8\yr$ \citep{Harris96}. The central relaxation timescale of our adopted King models is roughly $9.0\times10^7\yr$ (with a half-mass relaxation timescale of $2.6\times10^8\yr$). These numbers indicate that there may be small changes in the stellar distribution over the course of the simulations, implying that static models provide an upper limit on the predicted ICM content. In $10^8\yr$, however, the total change in mass of the RGB and AGB populations due to stellar winds is only $2\%$, which is negligible. In our simulations, the effect of dynamical heating of the stellar distribution due to mass loss can therefore be ignored and the use of static models is appropriate.

\subsection{Summary}
\label{subsec:summary}

We validate a 3D hydrodynamics simulation and undertake numerical investigations on the effect of the ICM interaction with the Galactic halo medium. We find that the GC's motion through the Halo, is sufficient in limiting the build-up of gas within ``typical'' mass globular clusters (i.e. $10^5\Msun$), to levels that are below most observational limits. For massive GCs ($10^6\Msun$), however, we predict the steady build-up of a cool reservoir of ICM that should be readily observable. We present simulations that employ a multi-mass GC model and incorporates empirical mass-loss formulae to model the RGB and AGB stellar winds.  The use of these mass-loss formulae provides a way to consistently model the rate of gas injection into our GC ICM. The rates obtained from these formulae, show that previously assumed mass-loss rates were very high and result in blue HB stars. The implementation of a multi-mass GC model, where the giant branch stars reside deeper in the gravitational potential well, results in the retention of more ICM gas than previously predicted with King models. Finally, we present some preliminary work in which we model the GC as a discrete, orbiting stellar population.  Although presented in its exploratory stages, our low-resolution $10^5\Msun$ GC simulations show that this approach readily predicts an ICM mass that is within observational limits. This result indicates that assuming a continuous stellar density distribution may overestimate the predicted GC gas content by a significant fraction. We summarise our main results below:
\begin{itemize}
  \item moving GC through the Galactic halo medium reduces ICM\\ \indent \indent content:
  \begin{itemize}
    \item $10^5\Msun$: ICM stripped to within observational limits
    \item $10^6\Msun$: ICM mass increases steadily with time
  \end{itemize}
  \item the mass of retained ICM gas is sensitive to the specific \\ \indent \indent mass-loss ($\alpha$) parameter used:
  \begin{itemize}
    \item empirical mass-loss laws give lower $\alpha$ than that used in\\ \indent\indent previous investigations;
  \end{itemize}
  \item more ICM is retained in a multi-mass GC model than in a King\\ \indent \indent model GC;
  \item discretising the stellar population amplifies and increases\\ \indent \indent the influence of ram-pressure stripping.
\end{itemize}

\subsection*{Acknowledgements}

The authors express thanks to our referee Jacco Van Loon for constructive comments which vastly improved this manuscript and to Ewald M\"{u}ller for his insightful comments on numerical hydrodynamics. WP acknowledges the Science and Technology Facilities Council and the Daiwa Anglo-Japanese Foundation for financial support. WP thanks Junichiro Makino for providing the opportunity to complete this work at the NAOJ and engaging in enlightening conversations about GC dynamics. WP also thanks Shoko Jin for comments that helped improve the presentation of this paper. WP wishes to acknowledge use of the high performance computing facilities at the ARI. Images were rendered using {\small VAPOR} (http://www.vapor.ucar.edu/).

\bibliographystyle{mn2e}
\bibliography{refs}

\begin{thebibliography}{}

\bibitem[\protect\citeauthoryear{{Agertz}, {Moore}, {Stadel}, {Potter},
  {Miniati}, {Read}, {Mayer}, {Gawryszczak}, {Kravtsov}, {Nordlund}, {Pearce},
  {Quilis}, {Rudd}, {Springel}, {Stone}, {Tasker}, {Teyssier}, {Wadsley} \&
  {Walder}}{{Agertz} et~al.}{2007}]{Agertz_et_al07}
{Agertz} O.,  {Moore} B.,  {Stadel} J.,  {Potter} D.,  {Miniati} F.,  {Read}
  J.,  {Mayer} L.,  {Gawryszczak} A.,  {Kravtsov} A.,  {Nordlund} {\AA}.,
  {Pearce} F.,  {Quilis} V.,  {Rudd} D.,  {Springel} V.,  {Stone} J.,  {Tasker}
  E.,  {Teyssier} R.,  {Wadsley} J.,    {Walder} R.,  2007, \mnras, 380, 963

\bibitem[\protect\citeauthoryear{{Barmby}, {Boyer}, {Woodward}, {Gehrz}, {van
  Loon}, {Fazio}, {Marengo} \& {Polomski}}{{Barmby}
  et~al.}{2009}]{Barmby_et_al09}
{Barmby} P.,  {Boyer} M.~L.,  {Woodward} C.~E.,  {Gehrz} R.~D.,  {van Loon}
  J.~T.,  {Fazio} G.~G.,  {Marengo} M.,    {Polomski} E.,  2009, \aj, 137, 207

\bibitem[\protect\citeauthoryear{{Bates}, {Kemp} \& {Montgomery}}{{Bates}
  et~al.}{1993}]{Bates_et_al93}
{Bates} B.,  {Kemp} S.~N.,    {Montgomery} A.~S.,  1993, \aaps, 97, 937

\bibitem[\protect\citeauthoryear{{Bode} \& {Evans}}{{Bode} \&
  {Evans}}{2008}]{BodeEvans08}
{Bode} M.~F.,  {Evans} A.,  2008, {Classical Novae}, {Second} edn.
No.~43 in {Cambridge Astrophysics Series}, {Cambridge University Press}

\bibitem[\protect\citeauthoryear{{Boyer}, {McDonald}, {Loon}, {Woodward},
  {Gehrz}, {Evans} \& {Dupree}}{{Boyer} et~al.}{2008}]{Boyer_et_al08}
{Boyer} M.~L.,  {McDonald} I.,  {Loon} J.~T.,  {Woodward} C.~E.,  {Gehrz}
  R.~D.,  {Evans} A.,    {Dupree} A.~K.,  2008, \aj, 135, 1395

\bibitem[\protect\citeauthoryear{{Boyer}, {Woodward}, {van Loon}, {Gordon},
  {Evans}, {Gehrz}, {Helton} \& {Polomski}}{{Boyer}
  et~al.}{2006}]{Boyer_et_al06}
{Boyer} M.~L.,  {Woodward} C.~E.,  {van Loon} J.~T.,  {Gordon} K.~D.,  {Evans}
  A.,  {Gehrz} R.~D.,  {Helton} L.~A.,    {Polomski} E.~F.,  2006, \aj, 132,
  1415

\bibitem[\protect\citeauthoryear{{Burke}}{{Burke}}{1968}]{Burke68}
{Burke} J.~A.,  1968, \mnras, 140, 241

\bibitem[\protect\citeauthoryear{{Capuzzo Dolcetta}, {Di Matteo} \&
  {Miocchi}}{{Capuzzo Dolcetta} et~al.}{2005}]{CapuzzoDolcetta_et_al05}
{Capuzzo Dolcetta} R.,  {Di Matteo} P.,    {Miocchi} P.,  2005, \aj, 129, 1906

\bibitem[\protect\citeauthoryear{{Cohen}}{{Cohen}}{1976}]{Cohen76}
{Cohen} J.~G.,  1976, \apjl, 203, L127

\bibitem[\protect\citeauthoryear{{Cohen} \& {Malkan}}{{Cohen} \&
  {Malkan}}{1979}]{CohenMalkan79}
{Cohen} N.~L.,  {Malkan} M.~A.,  1979, \aj, 84, 74

\bibitem[\protect\citeauthoryear{{Colella} \& {Woodward}}{{Colella} \&
  {Woodward}}{1984}]{ColellaWoodward84}
{Colella} P.,  {Woodward} P.~R.,  1984, Journal of Computational Physics, 54,
  174

\bibitem[\protect\citeauthoryear{{Coleman} \& {Worden}}{{Coleman} \&
  {Worden}}{1977}]{ColemanWorden77}
{Coleman} G.~D.,  {Worden} S.~P.,  1977, \apj, 218, 792

\bibitem[\protect\citeauthoryear{{Conklin} \& {Kimble}}{{Conklin} \&
  {Kimble}}{1974}]{ConklinKimble74}
{Conklin} I.~K.,  {Kimble} R.~A.,  1974, in Bulletin of the American
  Astronomical Society Vol.~6 of Bulletin of the American Astronomical Society,
  {Upper Limits to Neutral Hydrogen in Four Globular Clusters.}.
p.~468

\bibitem[\protect\citeauthoryear{{Cox} \& {Daltabuit}}{{Cox} \&
  {Daltabuit}}{1971}]{CoxDaltabuit71}
{Cox} D.~P.,  {Daltabuit} E.,  1971, \apj, 167, 113

\bibitem[\protect\citeauthoryear{{Cox} \& {Tucker}}{{Cox} \&
  {Tucker}}{1969}]{CoxTucker69}
{Cox} D.~P.,  {Tucker} W.~H.,  1969, \apj, 157, 1157

\bibitem[\protect\citeauthoryear{{Da Costa} \& {Freeman}}{{Da Costa} \&
  {Freeman}}{1976}]{DaCostaFreeman76}
{Da Costa} G.~S.,  {Freeman} K.~C.,  1976, \apj, 206, 128

\bibitem[\protect\citeauthoryear{{Dobrotka}, {Lasota} \& {Menou}}{{Dobrotka}
  et~al.}{2006}]{Dobrotka_et_al06}
{Dobrotka} A.,  {Lasota} J.,    {Menou} K.,  2006, \apj, 640, 288

\bibitem[\protect\citeauthoryear{{Dupree}}{{Dupree}}{1986}]{Dupree86}
{Dupree} A.~K.,  1986, \araa, 24, 377

\bibitem[\protect\citeauthoryear{{Dupree}, {Smith} \& {Strader}}{{Dupree}
  et~al.}{2009}]{Dupree_et_al09}
{Dupree} A.~K.,  {Smith} G.~H.,    {Strader} J.,  2009, \aj, 138, 1485

\bibitem[\protect\citeauthoryear{{Evans}, {Stickel}, {van Loon}, {Eyres},
  {Hopwood} \& {Penny}}{{Evans} et~al.}{2003}]{Evans_et_al03}
{Evans} A.,  {Stickel} M.,  {van Loon} J.~T.,  {Eyres} S.~P.~S.,  {Hopwood}
  M.~E.~L.,    {Penny} A.~J.,  2003, \aap, 408, L9

\bibitem[\protect\citeauthoryear{{Faulkner} \& {Freeman}}{{Faulkner} \&
  {Freeman}}{1977}]{FF77}
{Faulkner} D.~J.,  {Freeman} K.~C.,  1977, \apj, 211, 77

\bibitem[\protect\citeauthoryear{{Frail} \& {Beasley}}{{Frail} \&
  {Beasley}}{1994}]{FrailBeasley94}
{Frail} D.~A.,  {Beasley} A.~J.,  1994, \aap, 290, 796

\bibitem[\protect\citeauthoryear{{Frank} \& {Gisler}}{{Frank} \&
  {Gisler}}{1976}]{FrankGisler76}
{Frank} J.,  {Gisler} G.,  1976, \mnras, 176, 533

\bibitem[\protect\citeauthoryear{{Freire}, {Kramer}, {Lyne}, {Camilo},
  {Manchester} \& {D'Amico}}{{Freire} et~al.}{2001}]{Freire_et_al01}
{Freire} P.~C.,  {Kramer} M.,  {Lyne} A.~G.,  {Camilo} F.,  {Manchester} R.~N.,
     {D'Amico} N.,  2001, \apjl, 557, L105

\bibitem[\protect\citeauthoryear{{Fukugita} \& {Peebles}}{{Fukugita} \&
  {Peebles}}{2006}]{FukugitaPeebles06}
{Fukugita} M.,  {Peebles} P.~J.~E.,  2006, \apj, 639, 590

\bibitem[\protect\citeauthoryear{{Gisler}}{{Gisler}}{1976}]{Gisler76}
{Gisler} G.~R.,  1976, \aap, 51, 137

\bibitem[\protect\citeauthoryear{{Gnedin} \& {Ostriker}}{{Gnedin} \&
  {Ostriker}}{1997}]{GnedinOstriker97}
{Gnedin} O.~Y.,  {Ostriker} J.~P.,  1997, \apj, 474, 223

\bibitem[\protect\citeauthoryear{{Harris}}{{Harris}}{1996}]{Harris96}
{Harris} W.~E.,  1996, \aj, 112, 1487

\bibitem[\protect\citeauthoryear{{Heiles} \& {Henry}}{{Heiles} \&
  {Henry}}{1966}]{HeilesHenry66}
{Heiles} C.,  {Henry} R.~C.,  1966, \apj, 146, 953

\bibitem[\protect\citeauthoryear{{Hopwood}, {Evans}, {Penny} \&
  {Eyres}}{{Hopwood} et~al.}{1998}]{Hopwood_et_al98}
{Hopwood} M.~E.~L.,  {Evans} A.,  {Penny} A.,    {Eyres} S.~P.~S.,  1998,
  \mnras, 301, L30

\bibitem[\protect\citeauthoryear{{Hopwood}, {Eyres}, {Evans}, {Penny} \&
  {Odenkirchen}}{{Hopwood} et~al.}{1999}]{Hopwood_et_al99}
{Hopwood} M.~E.~L.,  {Eyres} S.~P.~S.,  {Evans} A.,  {Penny} A.,
  {Odenkirchen} M.,  1999, \aap, 350, 49

\bibitem[\protect\citeauthoryear{{Kerr}, {Bowers} \& {Knapp}}{{Kerr}
  et~al.}{1976}]{Kerr_et_al76}
{Kerr} F.~J.,  {Bowers} P.~F.,    {Knapp} G.~R.,  1976, in Bulletin of the
  American Astronomical Society Vol.~8 of Bulletin of the American Astronomical
  Society, {A Search for HI and OH in Southern Globular Clusters.}.
p.~537

\bibitem[\protect\citeauthoryear{{Kerr} \& {Knapp}}{{Kerr} \&
  {Knapp}}{1972}]{KerrKnapp72}
{Kerr} F.~J.,  {Knapp} G.~R.,  1972, \aj, 77, 573

\bibitem[\protect\citeauthoryear{{King}}{{King}}{1966}]{King66}
{King} I.~R.,  1966, \aj, 71, 64

\bibitem[\protect\citeauthoryear{{Knapp}, {Gunn}, {Bowers} \& {Vasquez
  Por\'{i}tz}}{{Knapp} et~al.}{1996}]{Knapp_et_al96}
{Knapp} G.~R.,  {Gunn} J.~E.,  {Bowers} P.~F.,    {Vasquez Por\'{i}tz} J.~F.,
  1996, \apj, 462, 231

\bibitem[\protect\citeauthoryear{{Knapp}, {Gunn} \& {Connolly}}{{Knapp}
  et~al.}{1995}]{Knapp_et_al95}
{Knapp} G.~R.,  {Gunn} J.~E.,    {Connolly} A.~J.,  1995, \apj, 448, 195

\bibitem[\protect\citeauthoryear{{Knapp} \& {Kerr}}{{Knapp} \&
  {Kerr}}{1973}]{KnappKerr73}
{Knapp} G.~R.,  {Kerr} F.~J.,  1973, \aj, 78, 458

\bibitem[\protect\citeauthoryear{{Knapp}, {Rose} \& {Kerr}}{{Knapp}
  et~al.}{1973}]{Knapp_et_al73}
{Knapp} G.~R.,  {Rose} W.~K.,    {Kerr} F.~J.,  1973, \apj, 186, 831

\bibitem[\protect\citeauthoryear{{Kroupa}}{{Kroupa}}{2002}]{Kroupa02}
{Kroupa} P.,  2002, Science, 295, 82

\bibitem[\protect\citeauthoryear{{Leon} \& {Combes}}{{Leon} \&
  {Combes}}{1996}]{LeonCombes96}
{Leon} S.,  {Combes} F.,  1996, \aap, 309, 123

\bibitem[\protect\citeauthoryear{{Maccarone} \& {Knigge}}{{Maccarone} \&
  {Knigge}}{2007}]{MaccaroneKnigge07}
{Maccarone} T.,  {Knigge} C.,  2007, Astronomy and Geophysics, 48, 050000

\bibitem[\protect\citeauthoryear{{Matsunaga}, {Mito}, {Nakada}, {Fukushi},
  {Tanab{\'e}}, {Ita}, {Izumiura}, {Matsuura}, {Ueta} \&
  {Yamamura}}{{Matsunaga} et~al.}{2008}]{Matsunaga_et_al08}
{Matsunaga} N.,  {Mito} H.,  {Nakada} Y.,  {Fukushi} H.,  {Tanab{\'e}} T.,
  {Ita} Y.,  {Izumiura} H.,  {Matsuura} M.,  {Ueta} T.,    {Yamamura} I.,
  2008, \pasj, 60, 415

\bibitem[\protect\citeauthoryear{{McDonald} \& {van Loon}}{{McDonald} \& {van
  Loon}}{2007}]{McDonaldVanLoon07}
{McDonald} I.,  {van Loon} J.~T.,  2007, \aap, 476, 1261

\bibitem[\protect\citeauthoryear{{McDonald}, {van Loon}, {Decin}, {Boyer},
  {Dupree}, {Evans}, {Gehrz} \& {Woodward}}{{McDonald}
  et~al.}{2009}]{McDonald_et_al09}
{McDonald} I.,  {van Loon} J.~T.,  {Decin} L.,  {Boyer} M.~L.,  {Dupree} A.~K.,
   {Evans} A.,  {Gehrz} R.~D.,    {Woodward} C.~E.,  2009, \mnras, 394, 831

\bibitem[\protect\citeauthoryear{{M{\'e}sz{\'a}ros}, {Avrett} \&
  {Dupree}}{{M{\'e}sz{\'a}ros} et~al.}{2009}]{Meszaros_et_al09}
{M{\'e}sz{\'a}ros} S.,  {Avrett} E.~H.,    {Dupree} A.~K.,  2009, \aj, 138, 615

\bibitem[\protect\citeauthoryear{{Miocchi}}{{Miocchi}}{2010}]{Miocchi10}
{Miocchi} P.,  2010, \aap, 514, A52

\bibitem[\protect\citeauthoryear{{O'Brien}, {Bode}, {Porcas}, {Muxlow},
  {Eyres}, {Beswick}, {Garrington}, {Davis} \& {Evans}}{{O'Brien}
  et~al.}{2006}]{OBrien_et_al06}
{O'Brien} T.~J.,  {Bode} M.~F.,  {Porcas} R.~W.,  {Muxlow} T.~W.~B.,  {Eyres}
  S.~P.~S.,  {Beswick} R.~J.,  {Garrington} S.~T.,  {Davis} R.~J.,    {Evans}
  A.,  2006, \nat, 442, 279

\bibitem[\protect\citeauthoryear{{Odenkirchen}, {Brosche}, {Geffert} \&
  {Tucholke}}{{Odenkirchen} et~al.}{1997}]{Odenkirchen_et_al97}
{Odenkirchen} M.,  {Brosche} P.,  {Geffert} M.,    {Tucholke} H.-J.,  1997, New
  Astronomy, 2, 477

\bibitem[\protect\citeauthoryear{{Okada}, {Kokubun}, {Yuasa} \&
  {Makishima}}{{Okada} et~al.}{2007}]{Okada_et_al07}
{Okada} Y.,  {Kokubun} M.,  {Yuasa} T.,    {Makishima} K.,  2007, \pasj, 59,
  727

\bibitem[\protect\citeauthoryear{{Origlia}, {Ferraro}, {Fusi Pecci} \&
  {Rood}}{{Origlia} et~al.}{2002}]{Origlia_et_al02}
{Origlia} L.,  {Ferraro} F.~R.,  {Fusi Pecci} F.,    {Rood} R.~T.,  2002, \apj,
  571, 458

\bibitem[\protect\citeauthoryear{{Origlia}, {Rood}, {Fabbri}, {Ferraro}, {Fusi
  Pecci} \& {Rich}}{{Origlia} et~al.}{2007}]{Origlia_et_al07}
{Origlia} L.,  {Rood} R.~T.,  {Fabbri} S.,  {Ferraro} F.~R.,  {Fusi Pecci} F.,
    {Rich} R.~M.,  2007, \apjl, 667, L85

\bibitem[\protect\citeauthoryear{{Origlia}, {Scaltriti}, {Anderlucci},
  {Ferraro} \& {Fusi Pecci}}{{Origlia} et~al.}{1997}]{Origlia_et_al97}
{Origlia} L.,  {Scaltriti} F.,  {Anderlucci} E.,  {Ferraro} F.~R.,    {Fusi
  Pecci} F.,  1997, \mnras, 292, 753

\bibitem[\protect\citeauthoryear{{Pietrinferni}, {Cassisi}, {Salaris} \&
  {Castelli}}{{Pietrinferni} et~al.}{2004}]{Pietrinferni_et_al04}
{Pietrinferni} A.,  {Cassisi} S.,  {Salaris} M.,    {Castelli} F.,  2004, \apj,
  612, 168

\bibitem[\protect\citeauthoryear{{Ramdani} \& {Jorissen}}{{Ramdani} \&
  {Jorissen}}{2001}]{RamdaniJorissen01}
{Ramdani} A.,  {Jorissen} A.,  2001, \aap, 372, 85

\bibitem[\protect\citeauthoryear{{Reimers}}{{Reimers}}{1975}]{Reimers75}
{Reimers} D.,  1975, M\'{e}moires of the Soci\'{e}t\'{e} Royale des Sciences de
  Li\`{e}ge, 8, 369

\bibitem[\protect\citeauthoryear{{Roberts}}{{Roberts}}{1988}]{Roberts88}
{Roberts} M.~S.,  1988, in {Grindlay} J.~E.,  {Philip} A.~G.~D.,  eds, The
  Harlow-Shapley Symposium on Globular Cluster Systems in Galaxies Vol.~126 of
  IAU Symposium, {Interstellar matter in globular clusters}.
pp 411--421

\bibitem[\protect\citeauthoryear{{Ruffert}}{{Ruffert}}{1992}]{Ruffert92}
{Ruffert} M.,  1992, \aap, 265, 82

\bibitem[\protect\citeauthoryear{{Scott} \& {Durisen}}{{Scott} \&
  {Durisen}}{1978}]{ScottDurisen78}
{Scott} E.~H.,  {Durisen} R.~H.,  1978, \apj, 222, 612

\bibitem[\protect\citeauthoryear{{Scott} \& {Rose}}{{Scott} \&
  {Rose}}{1975}]{ScottRose75}
{Scott} E.~H.,  {Rose} W.~K.,  1975, \apj, 197, 147

\bibitem[\protect\citeauthoryear{{Silk}}{{Silk}}{1974}]{Silk74}
{Silk} J.,  1974, Comments on Astrophysics and Space Physics, 6, 1

\bibitem[\protect\citeauthoryear{{Smith}, {Wood}, {Faulkner} \&
  {Wright}}{{Smith} et~al.}{1990}]{Smith_et_al90}
{Smith} G.~H.,  {Wood} P.~R.,  {Faulkner} D.~J.,    {Wright} A.~E.,  1990,
  \apj, 353, 168

\bibitem[\protect\citeauthoryear{{Smith}, {Woodsworth} \& {Hesser}}{{Smith}
  et~al.}{1995}]{Smith_et_al95}
{Smith} G.~H.,  {Woodsworth} A.~W.,    {Hesser} J.~E.,  1995, \mnras, 273, 632

\bibitem[\protect\citeauthoryear{{Smith}, {Hesser} \& {Shawl}}{{Smith}
  et~al.}{1976}]{smith_et_al76}
{Smith} M.~G.,  {Hesser} J.~E.,    {Shawl} S.~J.,  1976, \apj, 206, 66

\bibitem[\protect\citeauthoryear{{Spergel}}{{Spergel}}{1991}]{Spergel91}
{Spergel} D.~N.,  1991, \nat, 352, 221

\bibitem[\protect\citeauthoryear{{Spitzer}}{{Spitzer}}{1956}]{Spitzer56}
{Spitzer} L.~J.,  1956, \apj, 124, 20

\bibitem[\protect\citeauthoryear{{Tayler} \& {Wood}}{{Tayler} \&
  {Wood}}{1975}]{TaylerWood75}
{Tayler} R.~J.,  {Wood} P.~R.,  1975, \mnras, 171, 467

\bibitem[\protect\citeauthoryear{{Umbreit}, {Chatterjee} \& {Rasio}}{{Umbreit}
  et~al.}{2008}]{Umbreit_et_al08}
{Umbreit} S.,  {Chatterjee} S.,    {Rasio} F.~A.,  2008, \apjl, 680, L113

\bibitem[\protect\citeauthoryear{{van Loon}, {Stanimirovi{\'c}}, {Evans} \&
  {Muller}}{{van Loon} et~al.}{2006}]{vanLoon_et_al06}
{van Loon} J.~T.,  {Stanimirovi{\'c}} S.,  {Evans} A.,    {Muller} E.,  2006,
  \mnras, 365, 1277

\bibitem[\protect\citeauthoryear{{van Loon}, {Stanimirovi{\'c}}, {Putman},
  {Peek}, {Gibson}, {Douglas} \& {Korpela}}{{van Loon}
  et~al.}{2009}]{vanLoon_et_al09}
{van Loon} J.~T.,  {Stanimirovi{\'c}} S.,  {Putman} M.~E.,  {Peek} J.~E.~G.,
  {Gibson} S.~J.,  {Douglas} K.~A.,    {Korpela} E.~J.,  2009, \mnras, 396,
  1096

\bibitem[\protect\citeauthoryear{{Vandenberg}}{{Vandenberg}}{1978}]{Vandenberg%
78}
{Vandenberg} D.~A.,  1978, \apj, 224, 394

\bibitem[\protect\citeauthoryear{{Vandenberg} \& {Faulkner}}{{Vandenberg} \&
  {Faulkner}}{1977}]{VF77}
{Vandenberg} D.~A.,  {Faulkner} D.~J.,  1977, \apj, 218, 415

\bibitem[\protect\citeauthoryear{{Vassiliadis} \& {Wood}}{{Vassiliadis} \&
  {Wood}}{1993}]{VassiliadisWood93}
{Vassiliadis} E.,  {Wood} P.~R.,  1993, \apj, 413, 641

\bibitem[\protect\citeauthoryear{{Yokoo} \& {Fukue}}{{Yokoo} \&
  {Fukue}}{1992}]{YokooFukue92}
{Yokoo} T.,  {Fukue} J.,  1992, \pasj, 44, L253

\end{thebibliography}

\end{document}